\documentclass[sigplan,10pt]{acmart}

\acmConference[PL'18]{ACM SIGPLAN Conference on Programming Languages}{January 01--03, 2018}{New York, NY, USA}
\acmYear{2018}
\acmISBN{} 
\acmDOI{} 
\startPage{1}

\setcopyright{none}

\usepackage{xspace} 
\usepackage{amsmath,amssymb,amsthm}
\usepackage{mathpartir}
\usepackage{stmaryrd}
\usepackage{soul}
\usepackage{listings}
\usepackage{rotating}
\usepackage{setspace} 
\usepackage{wasysym}
\usepackage{thmtools}
\usepackage{mdframed}
\usepackage{xcolor}
 \usepackage{cleveref} 
\usepackage{url}
\usepackage{float}
\usepackage[utf8]{inputenc}
\usepackage{microtype}
\usepackage{enumitem}

\usepackage{mathtools}
\DeclarePairedDelimiter{\ceil}{\lceil}{\rceil}
\DeclarePairedDelimiter{\floor}{\lfloor}{\rfloor}
\usepackage{xargs}

\newif\ifextra

\newcommand{\tname}{\textsf{RelCost}\xspace}
\newcommand{\bitname}{\textsf{BiRelCost}\xspace}
\newcommand{\tnamemin}{\textsf{RelCostCore}}

\newcommand{\relstlc}{\textsf{relSTLC}\xspace}
\newcommand{\relref}{\textsf{RelRef}\xspace}
\newcommand{\relrefcore}{\textsf{RelRef Core}\xspace}
\newcommand{\birelref}{\textsf{BiRelRef}\xspace}
\newcommand{\relinf}{\textsf{RelInf}\xspace}

\newcommand{\relinfref}{\textsf{RelRefU}\xspace}
\newcommand{\relinfrefcore}{\textsf{RelRefU Core}\xspace}
\newcommand{\birelinfref}{\textsf{BiRelRefU}\xspace}

\newcommand{\ie}{i.\,e.}

\newcommand{\ssum}[4]{\sum\limits_{#1=#2}^{#3}#4}
\newcommand{\smin}[2]{\kw{min}({#1},{#2})}
\newcommand{\smax}[2]{\kw{max}(#1,#2)}

\newcommand{\szero}{0}

\newcommand{\splus}[2]{#1 + #2}

\newcommand{\sminus}[2]{#1 - #2}
\newcommand{\sdiv}[2]{\frac{#1}{#2}}
\newcommand{\smult}[2]{#1\cdot#2}
\newcommand{\splusone}[1]{#1+1}
\newcommand{\sceil}[1]{\ceil*{#1}}
\newcommand{\sfloor}[1]{\floor*{#1}}

\newcommand{\slog}[1]{\kw{log}_2(#1)}

\newcommand{\ssize}{\mathbb{N}}

\newcommand{\scost}{\mathbb{R}}

\newcommand{\sort}{S}
\newcommand{\sorted}[1]{#1 \mathrel{::} \sort}
\newcommand{\sized}[1]{#1 \mathrel{::} \ssize}

\newcommand{\grt}{A}

\newcommand{\tbool}{\mbox{bool}}
\newcommand{\trbool}{\mbox{bool}_r}
\newcommand{\tubool}{\mbox{bool}_u}

\newcommand{\tint}{\mbox{int}}
\newcommand{\tunit}{\mbox{unit}}
\newcommand{\trunit}{\mbox{unit}_r}
\newcommand{\tlist}[3]{\mbox{list}[#1]^{#2}\,#3}
\newcommand{\ulist}[2]{\mbox{list}[#1]\,#2}

\newcommand{\uarr}[2]{\mathrel{\xrightarrow[]{\wexec(#1,#2)}}}

\newcommand{\tarrd}[1]{\mathrel{\xrightarrow{\wdiff(#1)}}}

\newcommand{\tforalld}[2]{\forall#1\overset{\wdiff(#2)}{::}S.\,}

\newcommand{\tsforall}[1]{\forall#1{::}S.\,}
\newcommand{\tforallN}[1]{\forall#1{::}\ssize.\,}
\newcommand{\texists}[1]{\exists#1{::}S.\,}
\newcommand{\texistsN}[1]{\exists#1{::}\ssize.\,}
\newcommand{\tcimpl}[2]{#1 \mathrel{\supset} #2}
\newcommand{\tcprod}[2]{#1 \mathrel{\&} #2}

\newcommand{\ttimes}{\mathrel{\times}}

\newcommand{\tinter}{\mathrel{\wedge}}

\newcommand{\tch}[2]{U\,(#1,#2)} 
 
\newcommand{\tcho}[1]{U\,#1}

\newcommand{\trm}[2]{|#1|_{#2}}
\newcommand{\trmo}[1]{|#1|}

\newcommand{\tbox}[1]{\square\,#1}

\newcommand{\tpushd}[1]{(#1)^{\downarrow\square}}

\newcommand{\eapp}{\;}

\newcommand{\etrue}{\mbox{true}}
 \newcommand{\efalse}{\mbox{false}}
\newcommand{\eif}{\mbox{if\;}} 
\newcommand{\ethen}{\mbox{\;then\;}}
\newcommand{\eelse}{\mbox{\;else\;}}

\newcommand{\elet}{\mbox{let\;}}
\newcommand{\clet}{\mbox{clet}\;}

\newcommand{\ein}{\mbox{\;in\;}}
\newcommand{\ecase}{\mbox{\;case\;}} 
\newcommand{\eof}{\mbox{\;of\;}}
\newcommand{\eas}{\mbox{\;as\;}} 
\newcommand{\ecelim}{\mbox{celim\;}}
\newcommand{\enil}{\mbox{nil}} 
\newcommand{\epack}{\mbox{pack\;}}
\newcommand{\eunpack}{\mbox{unpack\;}}
\newcommand{\efix}{\mbox{fix\;}} 
 
\newcommand{\eLam}{ \Lambda}
\newcommand{\elam}{ \lambda} 
\newcommand{\eApp}{ [\,]\,}
 
\newcommand{\ewith}{\;\mbox{with}\;} 

\newcommand{\econs}{\mbox{cons}} 
\newcommand{\econsC}{\mbox{cons$_C$}} 
\newcommand{\econsNC}{\mbox{cons$_{NC}$}} 

\newcommand{\eswitch}{\kw{switch}\;}
\newcommand{\enoch}{\kw{NC}\;}
\newcommand{\eder}{\kw{der}\;}
\newcommand{\esplit}{\kw{split}\;}
\newcommand{\ecoerce}[2]{\kw{coerce}_{#1,#2}\;}

\newcommand{\caseof}[1]{\mbox{case}~#1~\mbox{of}}
\newcommand{\ofnil}[1]{~~\mbox{nil}~\to#1}

\newcommand{\ofcons}[3]{|~#1::#2~\to~#3}

\newcommand{\wdiff}{\mbox{\scriptsize diff}}
\newcommand{\wexec}{\mbox{\scriptsize exec}}

\newcommand{\rdiff}{\ominus}

\newcommand{\ldiff}{\lesssim}

\newcommand{\dd}[1]{\mathcal{D}\llbracket\Delta\rrbracket}

\newcommand{\jtype}[4]{\mathrel{\vdash_{#1}^{#2} {#3} : {#4}}}
\newcommand{\jtypeM}[4]{\mathrel{\vdash_{#1}^{#2} {#3} :^c {#4}}}

\newcommand{\jstype}[3]{\mathrel{\vdash
    {#1} \backsim {#2} : {#3}}}

\newcommand{\jtypediff}[4]{\mathrel{\vdash
    {#2} \ominus {#3} \ldiff #1 : {#4}}}

\newcommand{\jtypediffM}[4]{\mathrel{\vdash
    {#2} \ominus {#3} \ldiff #1 :^c {#4}}}

\newcommand{\jelab}[6]{\mathrel{\vdash
    {#2} \ominus {#3} \rightsquigarrow {#4} \ominus {#5} \ldiff #1 : {#6}}}

\newcommand{\jelabun}[5]{\mathrel{\vdash_{#1}^{#2}
    {#3} \rightsquigarrow {#4} : {#5}}}

\newcommand{\jelabcu}[4]{\mathrel{\vdash_{#1}^{#2}
    {#3} \rightsquigarrow {#3}^* : {#4}}}

\newcommand{\sty}[2]{\vdash#1 \mathrel{::} #2}

\newcommand{\jsubtype}[2]{\sat#1\sqsubseteq#2}
\newcommand{\jasubtype}[2]{\sat^{\mathsf{\grt}}#1\sqsubseteq#2}
\newcommand{\jeqtype}[2]{\sat#1 \equiv#2}

\newcommand{\sat}[1]{\models#1}

\newcommand{\ceq}[2]{#1\mathrel{\doteq}#2}
\newcommand{\cleq}[2]{#1 \mathop{\leq} #2}

\newcommand{\ceqz}[1]{#1 \mathrel{\doteq} 0}

\newcommand{\cand}[2]{#1 \wedge #2}
\newcommand{\cexists}[3]{\exists#1::#2.#3}

\newcommand{\cexistsall}[2]{\exists(#1).#2}
\newcommand{\cforall}[3]{\forall#1::#2.#3}

\newcommand{\cimpl}[2]{#1\rightarrow#2}

\newcommand{\ctrue}{\top}

\newcommand{\wfty}[1]{\vdash   #1~\kw{wf}}

\newcommand{\wfcs}[1]{\vdash   #1~\kw{wf}}

\newcommand{\ctx}{\Delta; \Phi_a; \Gamma}

\newcommand{\octx}{\Delta; \Phi_a; \Omega}

\newcommand{\fiv}[1]{\text{FIV}(#1)}

\let\oldstar\star
\renewcommand{\star}{\oldstar}

\newcommand{\im}[1]{\ensuremath{#1}}

\newcommand{\kw}[1]{\im{\mathtt{#1}}}

\newcommand{\omitthis}[1]{}

\newenvironment{nop}{}{}

\newenvironment{stackAux}[2]{%
\setlength{\arraycolsep}{0pt}
\begin{array}[#1]{#2}}{
\end{array}}

\newtheoremstyle{athm}{\topsep}{\topsep}%
      {\upshape}
      {}
      {\bfseries}
      {}
      {.8em}
      {\thmname{#1}\thmnumber{ #2}\thmnote{~\,(#3)}
      {\addcontentsline{def}{section}{#1~#2~#3}}%
\newline}

 \theoremstyle{athm}

\newtheorem{thm}{Theorem}

 \newtheorem{lem}[thm]{Lemma}

\newcommand{\bnfalt}{{\bf \,\,\mid\,\,}}

\renewcommand{\theenumii}{\theenumi.\arabic{enumii}.}

\usepackage{enumitem}
\setenumerate{listparindent=\parindent}

\newlist{enumih}{enumerate}{3}
\setlist[enumih]{label=\alph*),before=\raggedright, topsep=1ex, parsep=0pt,  itemsep=1pt }

\newlist{enumconc}{enumerate}{3}
\setlist[enumconc]{leftmargin=0.5cm, label*= \arabic*.  , topsep=1ex, parsep=0pt,  itemsep=3pt }

\newlist{enumsub}{enumerate}{3}
\setlist[enumsub]{ leftmargin=0.7cm, label*= \textbf{subcase} \bf \arabic*: }

\newlist{enumsubsub}{enumerate}{3}
\setlist[enumsubsub]{ leftmargin=0.5cm, label*= \textbf{subsubcase} \bf \arabic*: }

\newlist{mainitem}{itemize}{3}
\setlist[mainitem]{ leftmargin=0cm , label= {\bf Case} }

\newenvironment{nstabbing}
  {\setlength{\topsep}{0pt}%
   \setlength{\partopsep}{0pt}%
   \tabbing}
  {\endtabbing} 
 
\newcommand{\freshSize}[1]{#1\in\text{fresh}(\ssize)}
\newcommand{\freshCost}[1]{#1\in\text{fresh}(\scost)}

\newcommand{\chdiff}[5]{\vdash{#1}\rdiff{#2}~{\downarrow}~#3,#4 \Rightarrow
{\color{red}#5}}

\newcommand{\chsdiff}[3]{\vdash{#1}\backsim{#2}~{\downarrow}~#3}

\newcommand{\infdiff}[6]{\vdash{#1}\rdiff{#2}~{\uparrow}~{\color{red}{#3}}\Rightarrow[{\color{red}#4}],{\color{red}#5},{\color{red}#6}}

\newcommand{\infsdiff}[3]{\vdash{#1}\backsim{#2}~{\uparrow}~{\color{red}{#3}}}

\newcommand{\chexec}[5]{\vdash{#1}~{\downarrow}~#2, #3, #4  \Rightarrow
{{\color{red}#5}}}

\newcommand{\infexec}[6]{\vdash{#1}~{\uparrow}~{\color{red}{#2}}\Rightarrow[{\color{red}#3}], {\color{red}#4},{\color{red}#5},{\color{red}#6}}

\newcommand{\emptypsi}{.}

\newcommand{\jalgeqtype}[3]{\sat#1\equiv#2\Rightarrow {\color{red}#3}}
\newcommand{\jalgasubtype}[3]{\sat^{\mathsf{\grt}}#1\sqsubseteq#2\Rightarrow {\color{red}#3}}

\newcommand{\jalgssubtype}[2]{\sat#1\leq#2}

\newcommand{\fivars}[1]{\text{FIV}(#1)}

\newcommand{\bctx}{\Delta; \psi_a; \Phi_a; \Gamma}

\newcommand{\suba}[1]{{#1}[\theta_a]}

\newcommand{\subta}[1]{{#1}[\theta\,\theta_a]}

\newcommand{\erty}[1]{|#1|}

\newcommandx{\tyswitch}[2][1=,2=]
{
    \inferrule{
    \Delta; \Phi_a; \trmo{\Gamma} \jtype{k_1}{t_1}{e_1}{\grt}~~#1\\\\
    \Delta; \Phi_a; \trmo{\Gamma} \jtype{k_2}{t_2}{e_2}{\grt}~~#2
    } {\ctx \jtypediff{t_1-k_2}{e_1}{e_2}{\tcho{\grt}}}~\textbf{switch}
}

\newcommandx{\tynochange}[2][1=,2=]
{
  \inferrule
{
\Delta; \Phi_a; \Gamma \jtypediff{t}{e}{e}{\tau}~~#1\\\\
\forall x \in dom(\Gamma).~~
\Delta; \Phi_a  \jsubtype{\Gamma(x)}{\tbox{\Gamma(x)}} ~~#2
}
{
\Delta; \Phi_a; \Gamma, \Gamma'; \Omega \jtypediff{0}{e}{e}{\tbox{\tau}}
} ~\textbf{nochange}
}

\newcommand{\tyrfix}
{  \inferrule{
       {\ifextra
       \Delta; \Phi_a \wfty{\tau_1 \tarrd{t} \tau_2}  
       \fi} \\
       \Delta; \Phi_a;  x: \tau_1, f : \tau_1 \tarrd{t} \tau_2, \Gamma  \jtypediff{t}{e_1}{e_2}{\tau_2}}
            {\ctx \jtypediff{0}{ \efix f(x).e_1}{\efix f(x).e_2}{
              \tau_1 \tarrd{t} \tau_2}}~\textbf{r-fix}}

\newcommandx{\tyrfixNC}[2][1=,2=]
{  \inferrule
  {
    {\ifextra \Delta; \Phi_a \wfty{\tau_1 \tarrd{t} \tau_2}
      \fi} \\
    \Delta; \Phi_a;  x: \tau_1, f : \tbox{(\tau_1 \tarrd{t} \tau_2)}, \Gamma  \jtypediff{t}{e}{e}{\tau_2}~~#1\\
    \forall x \in dom(\Gamma).~~
    \Delta; \Phi_a  \jsubtype{\Gamma(x)}{\tbox{\Gamma(x)}}~~#2
  }
  {\ctx \jtypediff{0}{ \efix f(x).e}{\efix f(x).e}{
              \tbox{(\tau_1 \tarrd{t} \tau_2)}}}~\textbf{r-fixNC}
}

\newcommand{\tyrappD}
{\inferrule{\ctx \jtypediff{t_1}{e_1}{e_1'}{\tau_1}
  \tarrd{t} \tau_2 \\\\
 \ctx \jtypediff{t_2}{ e_2}{e_2'}{\tau_1}}
{\ctx \jtypediff{t_1+t_2+t}{e_1 \eapp e_2}{e_1' \eapp e_2'}{\tau_2}}~\textbf{r-app}}

\newcommandx{\tyriLam}[2][1=,2=]
{ \inferrule{
    i::S, \ctx \jtypediff{t}{e}{e'}{\tau}~~#1 \\\\
    i \not \in \fiv{\Phi_a; \Gamma} 
  }
  { \ctx \jtypediff{0}{\eLam e}{\eLam e'}{\tforalld{i}{t} \tau}}~\textbf{r-iLam}}
\newcommandx{\tyriApp}[2][1=,2=]
{  \inferrule{
    \ctx \jtypediff{t}{e}{e'}{\tforalld{i}{t'} \tau} ~~#1\\\\
    \Delta \vdash I : S ~~#2
  }
  {\ctx \jtypediff{t+t'[I/i]}{e \eApp}{e' \eApp}{\tau\{ I/ i\}}}~\textbf{r-iApp}}

\newcommandx{\tyrpack}[2][1=,2=]
{
\inferrule
{
  \ctx  \jtypediff{t}{e}{e'}{\tau\{I/i\}}~~#1\\
  \Delta \sty{I}{\sort}~~#2
}
{
  \ctx  \jtypediff{t}{\epack e}{\epack e'}{\texists{i} \tau}
} ~\textbf{r-pack}
}

\newcommandx{\tyrunpackD}[2][1=,2=]
{
\inferrule
{
  \ctx \jtypediff{t_1}{e_1}{e_1'}{\texists{i} \tau_1} ~~#1\\
  \sorted{i}, \Delta; \Phi_a; x: \tau_1, \Gamma \jtypediff{t_2}{e_2}{e_2'}{\tau_2} ~~#2\\
  i \not\in FV(\Phi_a;\Gamma, \tau_2, t_2) \\
}
{
  \ctx  \jtypediff{t_1+t_2}{\eunpack e_1 \eas x \ein  e_2}{\eunpack e_1' \eas x \ein  e_2'}{\tau_2}
}~\textbf{r-unpack1}}

\newcommandx{\tyrcaseL}[4][1=,2=,3=,4=]
{ \inferrule
  {\ctx \jtypediff{t}{e}{e'}{\tlist{n}{\alpha}{\tau}} ~~#1\\
   \Delta; \Phi_a \wedge n = 0  ;\Gamma \jtypediff{t'}{e_1}{e_1'}{\tau'} ~~#2\\
   i, \Delta; \Phi_a \wedge n = i+1;  h: \tbox{\tau},
   tl : \tlist{i}{\alpha}{\tau}, \Gamma \jtypediff{t'}{e_2}{e_2'}{\tau'} ~~#3
\\
   i, \beta, \Delta; \Phi_a \wedge n = i+1 \wedge \alpha = \beta +1 ;
   h: \tau, tl : \tlist{i}{\beta}{\tau}, \Gamma \jtypediff{t'}{e_2}{e_2'}{\tau'} ~~#4
   }
  {\ctx \jtypediff{t+t'}{ \ecase e \eof \enil \rightarrow e_1 ~|~ h::tl \rightarrow
    e_2}{\ecase e' \eof \enil \rightarrow e_1' ~|~ h::tl \rightarrow
    e_2'}{\tau'}}~\textbf{r-caseL}
}

\newcommandx{\tyinterE}[1][1=i]
{
   \inferrule{
     \octx \jtype{k}{t}{e}{\grt_1 \tinter \grt_2} \\
}{
  \octx \jtype{k}{t}{e}{\grt_{#1}} }~\textbf{interE$_{#1}$}
}

\newcommandx{\tyrinterE}[1][1=i]
{
   \inferrule{
    \ctx \jtypediff{t}{e}{e'}{\tau_1 \tinter \tau_2}\\
}{
  \ctx \jtypediff{t}{ e}{e'}{\tau_{#1}} }~\textbf{r-interE$_{#1}$}
}

\newcommandx{\tysubsum}[4][1=,2=,3=,4=]
{
   \inferrule{
    \octx \jtype{k}{t}{e}{\grt} ~~#1\\
    \Delta; \Phi_a \jasubtype{\grt}{\grt'} ~~#2\\
    \Delta; \Phi_a \sat k' \leq k ~~#3\\
    \Delta; \Phi_a \sat t \leq t'~~#4
}{
  \octx \jtype{k'}{t'}{e}{\grt'}
}~{\pmb{\sqsubseteq}\wexec}
}

\newcommandx{\tyrsubsum}[3][1=,2=,3=]
{
   \inferrule{
    \ctx \jtypediff{t}{e_1}{e_2}{\tau}~~#1\\
    \Delta; \Phi_a \jsubtype{\tau}{\tau'}~~#2 \\\\
    \Delta; \Phi_a \sat t \leq t'~~#3
  }
  {\ctx \jtypediff{t'}{e_1}{e_2}{\tau'} }~{\textbf{r-}\pmb{\sqsubseteq}}
}

\newcommandx{\starrD}[2][1=, 2=]
{
  \inferrule{
\Delta; \Phi_a  \jsubtype{\tau_1'}{\tau_1}~~ #1 \\
\Delta;\Phi_a  \jsubtype{\tau_2}{\tau_2'}~~ #2 \\\\
\Delta; \Phi_a \sat t \leq t'
}
{
\Delta; \Phi_a \jsubtype{\tau_1 \tarrd{t} \tau_2}{\tau_1' \tarrd{t'} \tau_2'}
}~\textbf{$\to \wdiff$}
}

\newcommandx{\stforallD}[1][1=]
{
  \inferrule
{
\sorted{i}, \Delta; \Phi_a  \jsubtype{\tau}{\tau'} ~~ #1\\
\sorted{i}, \Delta; \Phi_a \sat \cleq{t}{t'} \\
 i \not\in FV(\Phi_a)
}
{
\Delta; \Phi_a \jsubtype{\tforalld{i}{t}{\tau}}{\tforalld{i}{t'}{\tau'}}
}~\textbf{$\forall \wdiff$}
}

\newcommandx{\sttimes}[2][1=, 2=]
{
  \inferrule{
\Delta; \Phi_a  \jsubtype{\tau_1}{\tau_1'} ~~#1\\
\Delta; \Phi_a  \jsubtype{\tau_2}{\tau_2'} ~~#2
}{
\Delta; \Phi_a  \jsubtype{\tau_1 \times \tau_2}{\tau_1' \times \tau_2'}
}~\textbf{$\times$}
}

\newcommandx{\stlist}[3][1=,2=,3=]
{
  \inferrule{
\Delta; \Phi_a  \sat \ceq{n}{n'} ~~#1 \\
\Delta; \Phi_a  \sat \cleq{\alpha}{\alpha'} ~~#2\\
\Delta; \Phi_a  \jsubtype{\tau}{\tau'}~~#3
}
{
\Delta; \Phi_a   \jsubtype{\tlist{n}{\alpha}{\tau}}{\tlist{n'}{\alpha'}{\tau'}}
}~\textbf{l1}
}

\newcommand{\stlistzero}
{
  \inferrule
{
\Delta; \Phi_a \models \ceqz{\alpha}
}
{
\Delta; \Phi_a  \jsubtype{\tlist{n}{\alpha}{\tau}}{\tlist{n}{\alpha}\tbox{{\tau}}}
}~\textbf{l2}
}

\newcommand{\stlistbox}
{
  \inferrule
{
}
{
\Delta; \Phi_a  \jsubtype{\tlist{n}{\alpha}{\tbox{\tau}}}{\tbox{(\tlist{n}{\alpha}{\tau})}}
}~\textbf{l$\square$}
}

\newcommandx{\stexists}[1][1=]{
  \inferrule
{
\sorted{i}, \Delta; \Phi_a  \jsubtype{\tau}{\tau'} ~~#1 \\
i \not\in FV(\Phi_a)
}
{
\Delta; \Phi_a  \jsubtype{\texists{i}{\tau}}{\texists{i}{\tau'}}
}~\textbf{$\exists$} 
}

\newcommandx{\stcimpl}[2][1=, 2=]
{
  \inferrule{
    \Delta; \Phi_a \wedge C' \sat C ~~#1\\
    \Delta; \Phi_a  \jsubtype{\tau}{\tau'} ~~#2
  }
  {
    \Delta; \Phi_a  \jsubtype{\tcimpl{C}{\tau}}{\tcimpl{C'}{\tau'}}
  }\textbf{c-impl}
}

\newcommandx{\stcprod}[2][1=,2=]
{
\inferrule{
\Delta; \Phi_a \wedge C \sat  C' ~~#1 \\
\Delta; \Phi_a  \jsubtype{\tau}{\tau'} ~~#2
}
{
\Delta; \Phi_a  \jsubtype{\tcprod{C}{\tau}}{\tcprod{C'}{\tau'}}
}\textbf{c-and}
}

\newcommand{\stbox}
{\inferrule
{
}
{
\Delta; \Phi_a  \jsubtype{\tbox{\tau}}{\tau}
}~\textbf{T}
}

\newcommandx{\stboxB}[1][1=]
{\inferrule
{
\Delta; \Phi_a  \jsubtype{\tau_1}{\tau_2} #1
}
{
\Delta; \Phi_a  \jsubtype{\tbox{\tau_1}}{\tbox{\tau_2}}
}~\textbf{B-$\square$}
}

\newcommandx{\sttrans}[2][1=,2=]
{
\inferrule{
\Delta; \Phi_a  \jsubtype{\tau_1}{\tau_2}~~#1 \\
\Delta; \Phi_a  \jsubtype{\tau_2}{\tau_3}~~#2
}
{
\Delta; \Phi_a  \jsubtype{\tau_1}{\tau_3}
}\textbf{trans}
}

\newcommandx{\mtyswitch}[2][1=,2=]
{
    \inferrule{
    \Delta; \Phi_a; \trmo{\Gamma} \jtypeM{k_1}{t_1}{e_1}{\grt}~~#1\\
    \Delta; \Phi_a; \trmo{\Gamma} \jtypeM{k_2}{t_2}{e_2}{\grt}~~#2
    } {\ctx \jtypediffM{t_1-k_2}{\eswitch e_1}{\eswitch e_2}{\tcho{\grt}}}~\textbf{c-switch}
}

\newcommandx{\mtyinterE}[1][1=i]
{
   \inferrule{
     \octx \jtypeM{k}{t}{e}{\grt_1 \tinter \grt_2} \\
}{
  \octx \jtypeM{k}{t}{e}{\grt_{#1}} }~\textbf{c-interE$_{#1}$}
}

\newcommandx{\mtyrinterE}[1][1=i]
{
   \inferrule{
    \ctx \jtypediffM{t}{e}{e'}{\tau_1 \tinter \tau_2}\\
}{
  \ctx \jtypediffM{t}{ e}{e'}{\tau_{#1}} }~\textbf{c-r-interE$_{#1}$}
}

\newcommandx{\chRswitch}[1][1=]
{
  \inferrule
  {
    \freshCost{k_1#1, t_1#1, k_2#1, t_2#1} \\
    \Delta; k_1#1, t_1#1, \psi_a; \Phi; \trmo{\Gamma} \chexec{e_1#1}{\grt}{k_1#1}{t_1#1}{\Phi_1#1}\\
    \Delta; k_2#1, t_2#1, \psi_a; \Phi; \trmo{\Gamma} \chexec{e_2#1}{\grt}{k_2#1}{t_2#1}{\Phi_2#1}
  }
  {
    \bctx  \chdiff{\eswitch e_1#1}{\eswitch e_2#1}{t}{\tcho{\grt}}{\cexists{k_1,t_1}{\scost}{(\cand{\Phi_1#1}{\cexists{k_2#1,t_2#1}{\scost}{\cand{\Phi_2#1}{\ceq{t_1#1-k_2#1}{t}}}
  })}}
   }~\textbf{alg-r-switch$\downarrow$}
}

\newcommand{\chRnochange}
{
  \inferrule
  {
    \freshCost{t'} \\
    \Delta; {t', \psi_a}; \Phi_a; \tbox{\Gamma}  \chdiff{e}{e}{\tau}{t'}{\Phi}
  }
  {
    \Delta; \psi_a; \Phi_a; \Gamma', \tbox{\Gamma} \chdiff{\enoch e}{\enoch e}{\tbox{\tau}}{t}{\cand{\ceq{0}{t}}{(\cexists{t'}{\scost}{\Phi})}}
  } ~\textbf{alg-r-nochange-$\downarrow$}
}

\newcommandx{\chRder}[2][1=\psi_a, 2=]
{
\inferrule
{
  \Delta; #1; \Phi_a; \Gamma \chdiff{e_1#2}{e_2#2}{\tbox{\tau}}{t}{\Phi}
}
{
  \Delta; #1; \Phi_a; \Gamma \chdiff{\eder e_1#2}{\eder e_2#2}{\tau}{t}{\Phi}
}~\textbf{alg-r-der-$\downarrow$} 
}

\newcommandx{\infvar}[1][1=\psi]{
\inferrule
{ 
\Omega(x) = \grt
}
{
\Delta; #1; \Phi_a; \Omega  \infexec{x}{\grt}{\emptypsi}{0}{0}{\ctrue}
} ~\textbf{alg-u-var-$\uparrow$}
}

\newcommand{\infRapp}
{
\inferrule
{
\bctx  \infdiff{e_1}{e_1'}{\tau_1 \tarrd{t_e} \tau_2}{\psi}{t_1}{\Phi_1} \\
\freshCost{t_2} \\
\Delta; {t_2,\psi, \psi_a}; \Phi_a; \Gamma \chdiff{e_2}{e_2'}{\tau_1}{t_2}{\Phi_2}
}
{
\bctx  \infdiff{e_1 \eapp e_2}{e_1' \eapp e_2'}{\tau_2}{t_2,\psi}{t_1+t_2+t_e}{\cand{\Phi_1}{\Phi_2}}
}~\textbf{alg-r-app-$\uparrow$}
}

\newcommand{\chRupdown}
{
\inferrule{
  \bctx \infdiff{e}{e'}{\tau'}{\psi}{t'}{\Phi_1}\\
  \Delta; {\psi, \psi_a}; \Phi_a \jalgeqtype{\tau'}{\tau}{\Phi_2} \\
}
{ 
  \bctx \chdiff{e}{e'}{\tau}{t}{\cexistsall{\psi}{\cand{\cand{\Phi_1}{\Phi_2}}{\cleq{t'}{t}}}}
} ~\textbf{alg-r-$\uparrow\downarrow$}
}

\newcommandx{\elswitch}[2][1=,2=]
{
    \inferrule{
    \Delta; \Phi_a; \trmo{\Gamma} \jelabun{k_1}{t_1}{e_1}{e_1^*}{\grt}~~#1\\
    \Delta; \Phi_a; \trmo{\Gamma} \jelabun{k_2}{t_2}{e_2}{e_2^*}{\grt}~~#2
    } {\ctx \jelab{t_1-k_2}{e_1}{e_2}{\eswitch e_1^*}{\eswitch e_2^*}{\tcho{\grt}}}~\textbf{e-switch}
}

\newcommandx{\elnochange}[2][1=,2=]
{
  \inferrule
  {
    \Delta; \Phi_a; \Gamma \jelab{t}{e}{e}{e^*}{e^*}{\tau}~~#1 \\
    \forall x_i \in dom(\Gamma), ~~e_i = \ecoerce{\Gamma(x_i)}{\tbox{\Gamma(x_i)}}~~#2
  }
  {
    \Delta; \Phi_a; \Gamma, \Gamma' \jelab{0}{e}{e}{\elet \overline{ y_i = e_i \eapp x_i} \ein \enoch e^* [\overline{y_i/x_i}]}{\elet \overline{ y_i = e_i \eapp x_i} \ein \enoch e^* [\overline{y_i/x_i}]}{\tbox{\tau}}
  } ~\textbf{e-nochange}
}

\newcommandx{\astboxB}[4][1=\tau_1, 2=\tau_2, 3=, 4= ]
{
\inferrule
{
  \Delta; \psi_a; \Phi_a \jalgeqtype{#1}{#2}{\Phi#4} ~~~#3
}
{ \Delta; \psi_a; \Phi_a \jalgeqtype{\tbox{#1}}{\tbox{#2}}{\Phi#4}
}\textbf{B-$\square$}
} 
 
\newcommandx{\astboxL}[4][1=\tau_1, 2=\tau_2, 3=, 4= ]
{
\inferrule
{
  \Delta; \psi_a; \Phi_a \jalgeqtype{#1}{#2}{\Phi#4} ~~~#3
}
{ \Delta; \psi_a; \Phi_a \jalgeqtype{\tbox{#1}}{#2}{\Phi#4}
}\textbf{L-$\square$}
}

\newcommandx{\astboxpushl}[4][1=\tau_1, 2=\tau_2, 3=, 4= ]
{
\inferrule
{
  \Delta; \psi_a; \Phi_a \jalgeqtype{\tpushd{#1}}{#2}{\Phi#4} ~~~ #3
}
{\Delta; \psi_a; \Phi_a \jalgeqtype{\tbox{#1}}{#2}{\Phi#4}
}\textbf{$\square$-push-l$_i$}
}

\newcommandx{\astboxpushr}[5][1=\tau_1, 2=\tau_2, 3=, 4=, 5= ]
{
\inferrule
{
  \Delta; \psi_a; \Phi_a \jalgeqtype{#1}{\tpushd{#2}}{\Phi#4} ~~ #3 \\
   #2 \not = \sigma_1 \tarrd{k} \sigma_2, #2 \not =  \tch{\grt_1}{\grt_2}~~#5
}
{\Delta; \psi_a; \Phi_a \jalgeqtype{#1}{\tbox{#2}}{\Phi#4}
}\textbf{$\square$-push-r$_i$}
}

\newcommandx{\astboxBR}[3][1=\tau_1, 2=\tau_2, 3= ]
{
\inferrule
{
  \Delta; \psi_a; \Phi_a \jalgeqtype{\tbox{#1}}{#2}{\Phi} ~~~ #3
}
{\Delta; \psi_a; \Phi_a \jalgeqtype{\tbox{#1}}{\tbox{#2}}{\Phi}
}\textbf{B-$\square$-R}
}

\newcommandx{\astunrel}[8][1=\grt_1, 2=\grt_2, 3=\grt_1', 4 =\grt_2', 5=, 6=, 7=,8=]
{
\inferrule
{
  \Delta; \psi_a; \Phi_a \jalgasubtype{#1}{#3}{\Phi_1#7} ~~#5 \\
  \Delta; \psi_a; \Phi_a \jalgasubtype{#3}{#1}{\Phi_1#7'} ~~#5  \\
  \Delta; \psi_a; \Phi_a \jalgasubtype{#2}{#4}{\Phi_2#8} ~~#6 \\
  \Delta; \psi_a; \Phi_a \jalgasubtype{#4}{#2}{\Phi_2#8'} ~~#6
}
{\Delta; \psi_a; \Phi_a \jalgeqtype{\tch{#1}{#2}}{\tch{#3}{#4}}{\cand{\cand{\Phi_1#7}{\Phi_1#7'}}{\cand{\Phi_2#8}{\Phi_2#8'}}}
}\textbf{U}
}

\newcommandx{\etboxB}[3][1=\tau, 2=\tau', 3=]
{
\inferrule
{
  \Delta; \Phi_a \jeqtype{#1}{#2} ~~~#3
}
{ \Delta; \Phi_a \jeqtype{\tbox{#1}}{\tbox{#2}}
}\textbf{eq-B-$\square$}
}

\newcommandx{\etunrel}[6][1=\grt_1, 2=\grt_2, 3=\grt_1', 4 =\grt_2', 5=, 6=]
{
\inferrule
{
  \Delta; \Phi_a \jasubtype{#1}{#3} ~~#5 \\
  \Delta; \Phi_a \jasubtype{#3}{#1} ~~#5  \\
  \Delta; \Phi_a \jasubtype{#2}{#4}~~#6 \\
  \Delta; \Phi_a \jasubtype{#4}{#2} ~~#6
}
{\Delta; \Phi_a \jeqtype{\tch{#1}{#2}}{\tch{#3}{#4}}
}\textbf{eq-U}
}

\newcommand{\jelabt}[6]{\mathrel{\vdash
    {#2} \backsim {#3} \rightsquigarrow {#4} \backsim {#5}  : {#6}}}

\newcommand{\jelabunt}[5]{\mathrel{\vdash
    {#3} \rightsquigarrow {#4} : {#5}}}

\newcommand{\jelabct}[4]{\mathrel{\vdash
    {#2} \backsim {#3} \rightsquigarrow {#2}^* \backsim {#3}^*  : {#4}}}

\newcommand{\chdifft}[5]{\vdash{#1}\rdiff{#2}~{\downarrow}~#3 \Rightarrow
{\color{red}#5}}

\newcommand{\infdifft}[6]{\vdash{#1}\rdiff{#2}~{\uparrow}~{\color{red}{#3}}\Rightarrow{\color{red}#6}}

\newcommand{\jstypet}[3]{\mathrel{\vdash
    {#1} \backsim {#2} :^c {#3}}}

\newcommand{\infexecT}[6]{\vdash{#1}~{\uparrow}~{\color{red}{#2}} \Rightarrow{\color{red}#6}}

\newcommand{\chexecT}[5]{\vdash{#1}~{\downarrow}~#2 \Rightarrow{\color{red}#5}  }

\newcommand{\jtypediffMt}[4]{\mathrel{\vdash
    {#2} \backsim {#3} :^c {#4}}}

\newcommand{\jtypeMt}[4]{\mathrel{\vdash {#3} :^c {#4}}}

\newcommand{\rstlcsub }{ \inferrule{
    \Gamma \jstype{e_1}{e_2}{\tau}\\
     \jsubtype{\tau}{\tau'}~~ 
  }
  {\Gamma \jstype{e_1}{e_2}{\tau'}
  }~{\textbf{r-$\sqsubseteq$}} }

\newcommandx{\relrefsubto}[2][1=, 2=]
{
  \inferrule{
\Delta; \Phi_a  \jsubtype{\tau_1'}{\tau_1}~~ #1 \\
\Delta;\Phi_a  \jsubtype{\tau_2}{\tau_2'}~~ #2 \\
}
{
\Delta; \Phi_a \jsubtype{\tau_1 \to \tau_2}{\tau_1' \to \tau_2'}
}~\textbf{$\to $}
}
\newcommandx{\relrefsubforallD}[1][1=]
{
  \inferrule
{
\sorted{i}, \Delta; \Phi_a  \jsubtype{\tau}{\tau'} ~~ #1\\
 i \not\in FV(\Phi_a)
}
{
\Delta; \Phi_a \jsubtype{\tsforall{i}{\tau}}{\tsforall{i}{\tau'}}
}~\textbf{$\forall \wdiff$}
}

\newcommandx{\relrefembedconsOne}[2][1=,2=,]
{\inferrule
  { \ctx \jelabct{t_1}{e_1}{e_1'}{\tau}~~#1 \\
    \ctx \jelabct{t_2}{e_2}{e_2'}{\tlist{n}{\alpha}{\tau}}~~#2
  }
  { \ctx \jelabt{t_1+t_2}{ \econs(e_1,e_2) }{\econs(e_1',e_2')}{
      \econsC(e_1^*,e_2^*)}{
      \econsC(e_1'^*,e_2'^*)}{\tlist{n+1}{\alpha+1}{\tau}}}~\textbf{e-r-cons1}}

\newcommandx{\relrefembednochange}[2][1=,2=,]{\inferrule
  {
    \Delta; \Phi_a; \Gamma \jelabt{t}{e}{e}{e^*}{e^*}{\tau}~~#1 \\
    \forall x_i \in dom(\Gamma), ~~e_i =
       \ecoerce{\Gamma(x_i)}{\tbox{\Gamma(x_i)}}~~#2\\
        \forall x_i \in dom(\Gamma), ~~\Delta; \Phi_a  \jsubtype{\Gamma(x_i)}{\tbox{\Gamma(x_i)}}
  }
  {
    \Delta; \Phi_a; \Gamma, \Gamma' \jelabt{0}{e}{e}{\elet \overline{
       y_i = e_i \eapp x_i} \ein \enoch e^* [\overline{\eder
       y_i/x_i}]}{\elet \overline{ y_i = e_i \eapp x_i} \ein \enoch
       e^* [\overline{\eder y_i/x_i}]}{\tbox{\tau}}
  } ~\textbf{e-nochange}}

\newcommandx{\relrefembedsub}[3][1=,2=,3=,] { \inferrule
  {
    \Delta; \Phi_a; \Gamma \jelabct{t}{e_1}{e_2}{\tau}~~#1 \\
    \Delta; \Phi_a \jsubtype{\tau}{\tau'} ~~#2\\
    e' = \ecoerce{\tau}{\tau'} ~~#3
  }
  {
    \Delta; \Phi_a; \Gamma \jelabt{t'}{e_1}{e_2}{e' \eapp e_1^*}{e' \eapp e_2^*}{\tau'} \\
  }~\textbf{e-r-$\sqsubseteq$}}

\newcommandx{\relreftync}[2][1=,2=,]
{  \inferrule
{
\Delta; \Phi_a; \Gamma \jstype{e}{e}{\tau}~~\\\\
\forall x \in dom(\Gamma).~~#1
\Delta; \Phi_a  \jsubtype{\Gamma(x)}{\tbox{\Gamma(x)}} ~~#2
}
{
\Delta; \Phi_a; \Gamma,\Gamma' \jstype{e}{e}{\tbox{\tau}}
} ~\textbf{rr-nochange}}

\newcommandx{\relreftysub}[2][1=,2=,]{
\inferrule{
    \ctx \jstype{e_1}{e_2}{\tau}~~#1\\
    \Delta; \Phi_a \jsubtype{\tau}{\tau'}~~ #2
  }
  {\ctx \jstype{e_1}{e_2}{\tau'}
  }~{\textbf{rr-$\sqsubseteq$}}}

\newcommandx{\relreftyfixnc}[2][1=,2=,]{  \inferrule
  {
    {\ifextra \Delta; \Phi_a \wfty{\tau_1 \to \tau_2}
      \fi} \\
    \Delta; \Phi_a;  x: \tau_1, f : \tbox{(\tau_1 \to \tau_2)}, \Gamma \jstype{e}{e}{\tau_2}~~#1\\
    \forall x \in dom(\Gamma).
    \Delta; \Phi_a  \jsubtype{\Gamma(x)}{\tbox{\Gamma(x)}} ~~#2
  }
  {\ctx \jstype{ \efix f(x).e}{\efix f(x).e}{
              \tbox{(\tau_1 \to \tau_2)}}}~\textbf{rr-fixNC}
}

\newcommandx{\relreftyilam}[2][1=,2=,]{
\inferrule{
    i::S, \ctx \jstype{e}{e'}{\tau}~~#1 \\\\
    i \not \in \fiv{\Phi_a; \Gamma}~~#2
  }
  { \ctx \jstype{\eLam e}{\eLam e'}{\tsforall{i} \tau }}~\textbf{rr-iLam}
}

\newcommandx{\relreftyiapp}[2][1=,2=,]{\inferrule{
    \ctx \jstype{e}{e'}{\tsforall{i} \tau } ~~#1\\\\
    \Delta \vdash I : S ~~#2
  }
  {\ctx \jstype{e \eApp}{e' \eApp}{\tau\{ I/
      i\}}}~\textbf{rr-iApp}
}

\newcommandx{\relreftypack}[2][1=,2=,]{\inferrule
{
  \ctx  \jstype{e}{e'}{\tau\{I/i\}}~~#1\\
  \Delta \sty{I}{\sort}~~#2
}
{
  \ctx  \jstype{\epack e}{\epack e'}{\texists{i} \tau}
} ~\textbf{rr-pack}
}

\newcommandx{\relreftyunpack}[2][1=,2=,]{\inferrule
{
  \ctx \jstype{e_1}{e_1'}{\texists{i} \tau_1} ~~#1\\
  \sorted{i}, \Delta; \Phi_a; x: \tau_1, \Gamma \jstype{e_2}{e_2'}{\tau_2} ~~#2\\
  i \not\in FV(\Phi_a;\Gamma, \tau_2, t_2) \\
}
{
  \ctx  \jstype{\eunpack e_1 \eas x \ein  e_2}{\eunpack e_1' \eas x \ein  e_2'}{\tau_2}
}~\textbf{rr-unpack1} 
}

\newcommandx{\relreftyapp}[2][1=,2=,]{
    \inferrule{
\ctx \jstype{e_1}{e_1'}{\tau_1
      \to \tau_2}~~#1 \\
      \ctx \jstype{ e_2}{e_2'}{\tau_1}~~#2 }
    {\ctx \jstype{e_1 \eapp e_2}{e_1' \eapp
        e_2'}{\tau_2}}~\textbf{rr-app}
}

\newcommandx{\relreftyfix}[1][1=,]{
\inferrule{
       \Delta; \Phi_a;  x: \tau_1, f : \tau_1 \to \tau_2, \Gamma   \jstype{e_1}{e_2}{\tau_2}~~#1
     }
     {\ctx \jstype{ \efix f(x).e_1}{\efix f(x).e_2}{
                \tau_1 \to \tau_2}}~\textbf{rr-fix}
}

\newcommandx{\relreftycimplI}[1][1=,]{
\inferrule
{
  \Delta; \Phi_a \wedge C; \Gamma \jstype{e}{e'}{\tau} #1
}
{
  \ctx  \jstype{e}{e'}{\tcimpl{C}{\tau}}
} ~\textbf{rr-c-impII}
}

\newcommandx{\relreftycaseL}[4][1=,2=,3=,4=,]{\inferrule
  {\ctx \jstype{e}{e'}{\tlist{n}{\alpha}{\tau}} ~~#1\\
   \Delta; \Phi_a \wedge n = 0  ;\Gamma \jstype{e_1}{e_1'}{\tau'} ~~#2\\
   i, \Delta; \Phi_a \wedge n = i+1;  h: \tbox{\tau},
   tl : \tlist{i}{\alpha}{\tau}, \Gamma \jstype{e_2}{e_2'}{\tau'} ~~#3
\\
   i, \beta, \Delta; \Phi_a \wedge n = i+1 \wedge \alpha = \beta +1 ;
   h: \tau, tl : \tlist{i}{\beta}{\tau}, \Gamma \jstype{e_2}{e_2'}{\tau'} ~~#4
   }
  {\ctx \jstype{ \ecase e \eof \enil \rightarrow e_1 ~|~ h::tl \rightarrow
    e_2}{\ecase e' \eof \enil \rightarrow e_1' ~|~ h::tl \rightarrow
    e_2'}{\tau'}}~\textbf{rr-caseL}}

\newcommandx{\relreftyconOne}[2][1=,2=,]{\inferrule
  { \ctx \jstype{e_1}{e_1'}{\tau}~~#1 \\
    \ctx \jstype{e_2}{e_2'}{\tlist{n}{\alpha}{\tau}}~~#2
  }
  { \ctx \jstype{ \econs(e_1,e_2) }{ \econs(e_1',e_2')}{\tlist{n+1}{\alpha+1}{\tau}}}~\textbf{rr-cons1}}

\newcommandx{\relreftyconTwo}[2][1=,2=,]{ \inferrule
  { \ctx \jstype{e_1}{e_1'}{\tbox{\tau}}~~#1 \\
    \ctx \jstype{e_2}{e_2'}{\tlist{n}{\alpha}{\tau}}~~#2 }
  { \ctx \jstype{ \econs(e_1,e_2) }{
      \econs(e_1',e_2')}{\tlist{n+1}{\alpha}{\tau}}}~\textbf{rr-cons2}}

\newcommandx{\relreftysplit}[2][1=,2=,]{\inferrule
  {
    \Delta; \Phi_a \wedge C; \Gamma  \jstype{e_1}{e_2}{\tau}~~#1\\
    \Delta; \Phi_a \wedge \neg C; \Gamma  \jstype{e_1}{e_2}{\tau}~~#2\\
    \Delta \wfcs{C}
  }
  {
    \ctx  \jstype{e_1}{e_2}{\tau} } ~\textbf{rr-split}}

\newcommandx{\relreftycprodI}[2][1=,2=,]{\inferrule
{
\Delta; \Phi_a \sat C~~#1 \\
  \Delta; \Phi_a \wedge C; \Gamma \jstype{e}{e'}{\tau}~~#2
}
{
  \ctx  \jstype{e}{e'}{\tcprod{C}{\tau}}
} ~\textbf{rr-c-andI}}

\newcommandx{\relreftycprodE}[2][1=,2=,]{\inferrule
{
  \ctx \jstype{e_1}{e_1 '}{\tcprod{C}{\tau_1}}\\
  \Delta; \Phi_a \wedge C; x :\tau_1, \Gamma   \jstype{ e_2}{e_2'}{\tau_2}~~#1
}
{
  \ctx  \jstype{\clet e_1 \eas x \ein e_2}{\clet e_1' \eas x \ein e_2'}{\tau_2}~~#2
} ~\textbf{rr-c-andE}}

\newcommandx{\relrefalgtycaseL}[4][1=,2=,3=,4=,]{
 \inferrule
{
\bctx \infdifft{e}{e'}{\tlist{n}{\alpha}{\tau}}{\psi}{t_1}{\Phi_e}~~#1\\
\Delta; {  \psi_a}; \cand{\ceqz{n}}{\Phi_a}; \Gamma  \chdifft{e_1}{e_1'}{\tau'}{t_2}{\Phi_1}~~#2 \\
\sized{i},\Delta; {\psi_a};  \cand{\ceq{n}{i+1}}{\Phi_a}; h :\tbox{\tau}, tl:\tlist{i}{\alpha}{\tau}, \Gamma \chdifft{e_2}{e_2'}{\tau'}{t_2}{\Phi_2}~~#3\\
\sized{i}, \sized{\beta},\Delta; { \psi_a};  \cand{\cand{\ceq{n}{i+1}}{\ceq{\alpha}{\beta + 1}}}{\Phi_a}; h :\tau, tl:\tlist{i}{\beta}{\tau}, \Gamma \chdifft{e_3}{e_3'}{\tau'}{t_2}{\Phi_3}~~#4\\
\Phi_{body} = {\cand{(\cimpl{\ceqz{n}}{\Phi_1})}{(\cforall{i}{\ssize}{\cimpl{(\ceq{n}{i+1})}{(\cand{\Phi_2}{\cforall{\beta}{\ssize}{\cimpl{(\ceq{\alpha}{\beta+1})}{\Phi_3}}})}})}}
}
{
\bctx \chdifft{\hspace{0em}
       \begin{minipage}[c]{10em}
         ~$\ecase e \eof \enil \, \rightarrow e_1$ \\
         $~~~~~~|~h\,::_{NC} tl\, \rightarrow e_2$ \\
         $~~~~~~|~h::_{C} tl~~~ \rightarrow e_3$
       \end{minipage}\hspace{0em}
     }{\hspace{0em}
       \begin{minipage}[c]{10em}
         ~ $\ecase e' \eof \enil \, \rightarrow e_1'$ \\
         $~~~~~~~~|~h\,::_{NC} tl\, \rightarrow e_2'$ \\
         $~~~~~~~~|~h::_{C} tl~~~ \rightarrow e_3'$
       \end{minipage}
     }{\tau'}{t}{{(\cand{\Phi_e}{\Phi_{body}})}}
}~\textbf{alg-r-caseL-$\downarrow$}}

\newcommandx{\chRdert}[2][1=\psi_a, 2=]
{
\inferrule
{
  \Delta; #1; \Phi_a; \Gamma \chdifft{e_1#2}{e_2#2}{\tbox{\tau}}{t}{\Phi}
}
{
  \Delta; #1; \Phi_a; \Gamma \chdifft{\eder e_1#2}{\eder e_2#2}{\tau}{t}{\Phi}
}~\textbf{alg-r-der-$\downarrow$}
}

\newcommandx{\relinfsubto}[2][1=, 2=]
{
  \inferrule{
 \jsubtype{\tau_1'}{\tau_1}~~ #1 \\
 \jsubtype{\tau_2}{\tau_2'}~~ #2 \\
}
{
\jsubtype{\tau_1 \to \tau_2}{\tau_1' \to \tau_2'}
}~\textbf{$\to $}
}

\newcommandx{\relinfsubtrans}[2][1=,2=]
{
\inferrule{
 \jsubtype{\tau_1}{\tau_2}~~#1 \\
  \jsubtype{\tau_2}{\tau_3}~~#2
}
{
\jsubtype{\tau_1}{\tau_3}
}\textbf{tran}
}

\newcommandx{\relinfcoreusub}[2][1=,2=,]
{
   \inferrule{
    \Omega \jtypeM{}{}{e}{\grt}~~#1 \\
    \jsubtype{\grt}{\grt'}~~#2 \\
}{
  \Omega \jtypeM{k'}{t'}{e}{\grt'}
}~{\textbf{c-u-$\sqsubseteq$}
}
}

\newcommandx{\relinfcoreswitch}[2][1=,2=,]{\inferrule{
    \trm{\Gamma}{1} \jtypeM{}{}{e_1}{\grt_1}~~#1\\
    \trm{\Gamma}{2} \jtypeM{}{}{e_2}{\grt_2}~~#2
    } {\Gamma \jstypet{\eswitch e_1}{\eswitch e_2}{\tch{\grt_1}{\grt_2}}}~\textbf{c-switch}}

\newcommandx{\relinfembedswitch}[2][1=,2=,]{\inferrule{
    \trm{\Gamma}{1} \jelabun{k_1}{t_1}{e_1}{e_1^*}{\grt_1}~~#1\\
    \trm{\Gamma}{2} \jelabun{k_2}{t_2}{e_2}{e_2^*}{\grt_2}~~#2
    } {\Gamma \jelabt{t_1-k_2}{e_1}{e_2}{\eswitch e_1^*}{\eswitch
    e_2^*}{\tch{\grt_1}{\grt_2}}}~\textbf{e-switch}}

\newcommandx{\relinfembedusub}[2][1=,2=,]
{
  \inferrule
  {
    \Omega \jelabcu{}{}{e}{\grt}~~#1 \\
     \jasubtype{\grt}{\grt'}~~#2 \\
  }
  {
    \Omega \jelabcu{}{}{e}{\grt'} \\
  }~\textbf{e-u-$\sqsubseteq$}
}

\newcommandx{\relrefinfmtyswitch}[2][1=,2=]
{
    \inferrule{
    \Delta; \Phi_a; \trm{\Gamma}{1} \jtypeMt{k_1}{t_1}{e_1}{\grt_1}~~#1\\
    \Delta; \Phi_a; \trm{\Gamma}{2} \jtypeMt{k_2}{t_2}{e_2}{\grt_2}~~#2
    } {\ctx \jtypediffMt{t_1-k_2}{\eswitch e_1}{\eswitch e_2}{\tch{\grt_1}{\grt_2}}}~\textbf{c-switch}
}

\newcommandx{\relrefinfmtyinterE}[1][1=i]
{
   \inferrule{
     \octx \jtypeMt{k}{t}{e}{\grt_1 \tinter \grt_2} \\
}{
  \octx \jtypeMt{k}{t}{e}{\grt_{#1}} }~\textbf{c-interE$_{#1}$}
}

\newcommandx{\relrefinfmtyrinterE}[1][1=i]
{
   \inferrule{
    \ctx \jtypediffMt{t}{e}{e'}{\tau_1 \tinter \tau_2}\\
}{
  \ctx \jtypediffMt{t}{ e}{e'}{\tau_{#1}} }~\textbf{c-r-interE$_{#1}$}
}

\newcommandx{\relrefinfelswitch}[2][1=,2=]
{
    \inferrule{
    \Delta; \Phi_a; \trm{\Gamma}{1} \jelabunt{}{}{e_1}{e_1^*}{\grt_1}~~#1\\
    \Delta; \Phi_a; \trm{\Gamma}{2} \jelabunt{}{}{e_2}{e_2^*}{\grt_2}~~#2
    } {\ctx \jelabt{t_1-k_2}{e_1}{e_2}{\eswitch e_1^*}{\eswitch e_2^*}{\tch{\grt_1}{\grt_2}}}~\textbf{e-switch}
}

\newcommandx{\relrefinfelnochange}[2][1=,2=]
{
  \inferrule
  {
    \Delta; \Phi_a; \Gamma \jelabt{t}{e}{e}{e^*}{e^*}{\tau}~~#1 \\
    \forall x_i \in dom(\Gamma), ~~e_i = \ecoerce{\Gamma(x_i)}{\tbox{\Gamma(x_i)}}~~#2
  }
  {
    \Delta; \Phi_a; \Gamma, \Gamma' \jelabt{0}{e}{e}{\elet \overline{
        y_i = e_i \eapp x_i} \ein \enoch e^* [\overline{\eder y_i/x_i}]}{\elet \overline{ y_i =\eder e_i \eapp x_i} \ein \enoch e^* [\overline{y_i/x_i}]}{\tbox{\tau}}
  } ~\textbf{e-nochange}
}

\newcommandx{\relrefinfelrsubsum}[3][1=,2=,3=,]
{
  \inferrule
  {
    \Delta; \Phi_a; \Gamma \jelabct{t}{e_1}{e_2}{\tau}~~#1 \\
    \Delta; \Phi_a \jsubtype{\tau}{\tau'} ~~#2\\
    e' = \ecoerce{\tau}{\tau'} ~~#3\\
  }
  {
    \Delta; \Phi_a; \Gamma \jelabt{t'}{e_1}{e_2}{e' \eapp e_1^*}{e' \eapp e_2^*}{\tau'} \\
  }~\textbf{e-r-$\sqsubseteq$}
}

\newcommandx{\rrfchRswitch}[1][1=]
{
  \inferrule
  {
    \Delta;  \psi_a; \Phi; \trm{\Gamma}{1}  \chexecT{e_1#1}{\grt_1}{k_1#1}{t_1#1}{\Phi_1#1}\\
    \Delta;  \psi_a; \Phi; \trm{\Gamma}{2}  \chexecT{e_2#1}{\grt_2}{k_2#1}{t_2#1}{\Phi_2#1}
  }
  {
    \bctx  \chdifft{\eswitch e_1#1}{\eswitch e_2#1}{t}{\tch{\grt_1}{\grt_2}}{\cand{\Phi_1#1}{\Phi_2#1}
  }
   }~\textbf{alg-r-switch$\downarrow$}
}

\newcommandx{\rrfchRder}[2][1=\psi_a, 2=]
{
\inferrule
{
  \Delta; #1; \Phi_a; \Gamma \chdifft{e_1#2}{e_2#2}{\tbox{\tau}}{t}{\Phi}
}
{
  \Delta; #1; \Phi_a; \Gamma \chdifft{\eder e_1#2}{\eder e_2#2}{\tau}{t}{\Phi}
}~\textbf{alg-r-der-$\downarrow$} 
}

\newcommandx{\rrfinfvar}[1][1=\psi]{
\inferrule
{ 
\Omega(x) = \grt
}
{
\Delta; #1; \Phi_a; \Omega  \infexecT{x}{\grt}{\emptypsi}{0}{0}{\ctrue}
} ~\textbf{alg-u-var-$\uparrow$}
}

\usepackage{booktabs}   
\usepackage{subcaption} 

\begin{document}

\title{Bidirectional Type Checking for Relational Properties}

\author{ Ezgi \c{C}i\c{c}ek } 
 \affiliation{ \institution{Facebook }  }
 \email{ezgicicekk@gmail.com}

\author{ Weihao Qu } 
 \affiliation{ \institution{ University at Buffalo, SUNY }  }
 \email{weihaoqu@buffalo.edu}

\author{ Gilles Barthe } 
 \affiliation{ \institution{IMDEA }  }
 \email{gilles.barthe@imdea.org}

\author{ Marco Gaboardi } 
 \affiliation{ \institution{ University at Buffalo, SUNY }  }
 \email{gaboardi@buffalo.edu}

\author{ Deepak Garg } 
 \affiliation{ \institution{MPI-SWS }  }
 \email{dg@mpi-sws.org}

\begin{abstract}
Relational type systems have been designed for several applications
including information flow, differential privacy, and cost analysis.
In order to achieve the best results, these systems often use
\emph{relational refinements} and \emph{relational effects} to
maximally exploit the similarity in the structure of the two programs
being compared. Relational type systems are appealing for relational
properties because they deliver simpler and more precise verification
than what could be derived from typing the two programs
separately. However, relational type systems do not yet achieve the
practical appeal of their non-relational counterpart, in part because
of the lack of a general foundations for implementing them.

In this paper, we take a step in this direction by developing
\emph{bidirectional relational} type checking for systems with
relational refinements and effects. Our approach achieves the benefits
of bidirectional type checking, in a relational setting. In
particular, it significantly reduces the need for typing annotations
through the combination of type checking and type inference.  In order
to highlight the foundational nature of our approach, we develop
bidirectional versions of several relational type systems which
incrementally combine many different components needed for expressive
relational analysis.
\end{abstract} 


\maketitle

\section{Introduction}
Type systems are a fundamental tool for proving program properties.
They draw their success from their ability to enforce many desirable
facts about programs. Bidirectional type checking~\citep{Pierce:2000}
is a quite recent but very successful method for implementing type
systems through a combination of type inference and type
checking~\cite{BrachaOSW98,ozz:colored-local,JonesVWS07,BiermanMT07,Abel17icfp}.
The appeal of bidirectional type checking lies in its ability to
minimize typing annotations---in most cases, type annotations are
needed only on recursive functions, or on reducible
expressions---while supporting disciplines that are too expressive to
fall under the purview of type inference. Furthermore, bidirectional
type systems offer a formal framework based on rules that resemble
standard typing rules. This simplifies proofs of soundness and
completeness of the algorithmic implementation relative to the
declarative type system.

Type systems are primarily focused on program properties, \ie\
reasoning about individual execution traces. In contrast, relational type
systems~\cite{AbadiCC93,Pottier03:IF,gaboardi13:dep,BartheFGSSB14,BGGHRS15,DuCostIt16,Cicek:2017,AguirreBG0S17,Aguirre18esop}
aim to repeat the success of type systems, but for so-called
relational properties, which consider pairs of execution
traces. Typical examples of relational properties include
non-interference in information flow systems, continuity and
robustness analysis of programs, differential privacy, and relational
cost analysis. The key difference of relational type systems is that
they consider two expressions simultaneously, and maximally exploit
structural similarities between them to achieve simpler and more
precise verification than would be possible with unary analysis of the
individual expressions. Similarities are exploited through two main
ingredients: relational refinement types and relational effects.

\emph{Relational refinements
  types}~\citep{Pottier03:IF,gaboardi13:dep,BGGHRS15,DuCostIt16,Cicek:2017}
relate two executions of two expressions and are akin to standard
refinement types~\cite{xi99:dependent}. However, their interpretation
is a relation between the values in the two executions. For example,
in information flow control, a relational refinement is used to
describe equivalence between the values that are observable at a
specific security level.

\emph{Relational
  effects}~\citep{Pottier03:IF,gaboardi13:dep,BGGHRS15,DuCostIt16,Cicek:2017}
are often of a quantitative nature and measure some quantitative
difference between two executions of the two expressions. These
relational effects are similar in spirit to their standard unary
counterpart~\cite{Lucassen:1988,nielsen99,PetricekOM13,BrunelGMZ14}
but their interpretation is a relation between the effects of the two
executions. For example, in differential privacy, a relational effect
is used to measure the level of indistinguishability between the
observable outputs on two inputs differing in one data element.

While several of the works cited above come with implemented type
checkers, there is, so far, no common understanding of the challenges
and solutions for implementing relational type systems. For this
reason, the broader goal of our work is to investigate issues in
\emph{implementing} a type checker for relational type systems with
relational refinements and relational effects. Bidirectional type
checking is a natural starting point for the reasons mentioned above,
and because it has been used for implementing refinement type
systems~\cite{xi99:dependent,Davies:2000,DunfieldK16} and
subtyping~\cite{Pierce:2000}, which are important common features in
most of the type systems we are inspired from. However, bidirectional
type checking has not been extensively applied to effect systems,
although some examples exist~\cite{Toro:2015}, and it has not been
applied to relational type systems.

\paragraph{Our contribution}
We present a study of \emph{bidirectional} type checking for
\emph{relational type and effect systems}. We start with the study of
a basic relational type system, named {\relstlc}, that includes
judgments to only relate two expressions with the same top-level
structure, with types to represent related and non-related boolean
values, no relational refinements and no relational effects. This can
be seen as the relational analogue of the simply typed lambda calculus
over a base type with subtyping. For this system, bidirectional type
checking works as expected and it delivers a sound and complete
algorithm implementing the declarative system.

Next, we extend {\relstlc} in two steps inspired by the features of
previously proposed relational type systems. Our first step, named
{\relref}, adds \emph{relational refinement types} over lists (as an
example of an inductive data type), and a comonadic type that
represents \emph{syntactic equality} of two values. Our second step,
named {\relinfref}, adds to {\relref} the possibility to \emph{relate
  arbitrary programs} of possibly dissimilar syntactic structure,
thanks to the possibility to switch to a complementary unary type
system.  Both these extensions add intrinsic \emph{nondeterminism} to
the type system to allow a programmer flexibility in writing
programs. The source of nondeterminism in both these systems is
non-syntax-directed typing and subtyping rules.  {\relref} has such
rules for relational refinement types and for subtyping, while
{\relinfref} has such a rule for switching to unary typing and more
such rules for subtyping.

To overcome the challenges introduced by nondeterminism, we introduce a
two-step methodology.  We first show that every well-typed program can
be translated to a well-typed program in a core language containing
term-level constructors that resolve the nondeterminism. This
translation is type derivation-directed; it introduces annotations to
resolve the nondeterminism in applying the (non-syntax-directed)
typing rules and does away with relational subtyping by replacing all
instances of relational subtyping with explicit coercions defined
within the core language.  Next, we develop a bidirectional type
system and prove it sound and complete with respect to the core
system. It follows that every typeable program can be annotated to
remove nondeterminism, and then bidirectionally type checked. This
proves the completeness of the bidirectional type checking modulo
nondeterminism.  We show that this methodology is applicable to both
{\relref} and {\relinfref}.

Our final step is to add \emph{relational effects} to
{\relinfref}. Specifically, we consider the type system
{\tname}~\cite{Cicek:2017}. This type system extends our {\relinfref}
with a relational effect to enable relational cost analysis. The
objective of relational cost analysis is to establish a static upper
bound on the cost of a program relative to another program: For two
programs $e_1$ and $e_2$, relational cost analysis establishes an
upper bound $t$ such that $cost(e_1) - cost(e_2) \leq t$. $t$ is
called the relative cost of $e_1$ and $e_2$. It is described as a
relational effect in the type system. Since {\tname} extends
{\relinfref}, it inherits the latter's many sources of
nondeterminism. We resolve these using the same two-step approach that
we described above, thus showing that the approach also extends to
relational effects.

To show the effectiveness of bidirectional type checking for
relational type systems, we have implemented a prototype for
{\tname}. (This prototype can also be used for the other type systems
we describe, since {\tname} extends them conservatively.)  Our
implementation handles the two steps of our approach
simultaneously. To implement the first step, rather than translating
type derivations to the core language, we use several example-guided
heuristics to resolve nondeterminism in applying the typing and
subtyping rules. We explain these heuristics and our evaluation shows
that they are effective for a large class of examples. For the second
step, we implement the bidirectional typing rules. Both type checking
and type inference generate constraints that capture arithmetic
relationships between refinements (e.g., list sizes) of various
subterms, relational refinements and relationships between unary and
relational costs. Our constraints contain existentially quantified
variables over integers and reals. Therefore, we design our own
algorithm to eliminate existential variables by finding substitutions
for them and use SMT solvers to discharge the constraints resulting
from substitutions.

Summing up, our contributions are:
\begin{itemize}
\item We present several bidirectional relational type systems that
  combine relational and non-relational typing, refinements, and
  unary and relational effects.
\item We present a type-preserving, complete embedding of programs
  that are typeable in those systems into core type systems. This embedding
  eliminates nondeterminism in applying typing rules and eliminates
  relational subtyping. We use the embedding to argue that, modulo the
  nondeterminism, our bidirectional type checking is complete.
\item We present an implementation of the largest of the type systems
  we consider ({\tname}), using heuristics to get rid of the inherent
  non-determinism. We use the implementation to type-check several
  examples, including all examples from the original {\tname} paper.
\end{itemize}
The rest of the paper is organized as follows. We start in
Section~\ref{sec:relstlc} with \relstlc, our basic relational
simply-typed calculus. In Sections~\ref{sec:relref}
and~\ref{sec:relinfref}, we extend it to {\relref} and
{\relinfref}. In Section~\ref{sec:relcost}, we add effects, finally
reaching {\tname}. In each of these sections, we describe a
declarative type system, its bidirectional version and, where
necessary, a core calculus that resolves nondeterminism of the
declarative type system. In Section~\ref{sec:implementation}, we
describe our implementation, heuristics to eliminate nondeterminism,
and experimental results. Section~\ref{sec:related} presents related
work. An anonymous appendix, provided as supplementary material for
the review, contains all technical details.

\section{Relational STLC (\relstlc)}
\label{sec:relstlc}

As an introduction to how relational reasoning works, we consider
{\relstlc}, a rehash of the simply-typed lambda calculus (STLC) with
relational reasoning. {\relstlc} has the following type and expression
grammar:
\[
\begin{array}{@{}ll@{}}
 \label{fig:syntax}
   \mbox{Types} 
  &  \tau  ::=  \trbool ~|~ \tubool ~|~  \tau_1 \to \tau_2  \\

  \mbox{Expr} & e  ::=  x  ~|~ \etrue ~|~ \efalse ~|~ \eif e
                                 \ethen e_1 \eelse e_2  ~|~ \elam x. e ~|~ e_1 \eapp e_2

\end{array}
\] 
A type $\tau$ is interpreted as a set of pairs of values. For
instance, the primitive type $\trbool$ ascribes pairs of
\emph{identical} booleans (the diagonal relation on booleans) whereas
the type $\tubool$ ascribes pairs of arbitrary booleans (the
complete relation on booleans). The function type $\tau_1 \to \tau_2$
relates pairs of functions that, given a pair of related arguments of
type $\tau_1$, return a pair of computations of type $\tau_2$. Even
though {\relstlc} is quite primitive, it forms the basis of our
development and we find it instructive to discuss challenges in its
algorithmization.

\paragraph{Declarative typing}
The typing judgment $\Gamma \jstype{e_1}{e_2}{\tau}$ ascribes the
expressions $e_1$ and $e_2$ the relational type $\tau$ under the
environment $\Gamma$. The typing rules and subtyping rules are
standard. A selection is shown
in~\Cref{fig:rel-stlc-typing-subtyping}. Note how the rule
\textbf{r-bool} relates two identical booleans at type $\trbool$,
while \textbf{r-u-bool} relates two arbitrary booleans at type
$\tubool$. This difference manifests in the rule \textbf{r-if}: If the
branch condition of an if-then-else has type $\trbool$, then we only
need to type the two ``then'' branches and the two ``else'' branches
separately, but do not need to type a ``then'' and an ``else'' branch
together. Finally, note that the calculus has (standard) subtyping
induced by the relation $\trbool \sqsubseteq \tubool$.





\begin{figure}

\raggedright
\fbox{  $\Gamma \jstype{e_1}{e_2}{\tau}$ } 
  \begin{mathpar}
    \inferrule
    { \kw{b}  \in \{\etrue, \efalse \}}  
    { \Gamma \jstype{\kw{b}}{\kw{b}}{\trbool}}~\textbf{r-bool}
    \and
    \inferrule
    { \kw{b_1, b_2}  \in \{\etrue, \efalse \}}  
    { \Gamma \jstype{\kw{b_1}}{\kw{b_2}}{\tubool}}~\textbf{r-u-bool}
    \and
    \inferrule{\Gamma \jstype{e}{e'}{\trbool} \\\\
      \Gamma \jstype{ e_1}{e_1'}{\tau} \\
      \Gamma \jstype{ e_2}{e_2'}{\tau}
    }
    {\Gamma \jstype{\eif e \ethen e_1 \eelse e_2}{\eif e' \ethen e_1' \eelse e_2'}{\tau}}~\textbf{r-if}
    \and
    \inferrule{
       \Gamma,  x: \tau_1 \jstype{e_1}{e_2}{\tau_2}
     }
     {\Gamma \jstype{ \elam x.e_1}{\elam x.e_2}{
                \tau_1 \to \tau_2}}~\textbf{r-lam}
    \and
    \inferrule{\Gamma \jstype{e_1}{e_1'}{\tau_1}
      \to \tau_2 \\
      \Gamma \jstype{ e_2}{e_2'}{\tau_1}}
    {\Gamma \jstype{e_1 \eapp e_2}{e_1' \eapp
        e_2'}{\tau_2}}~\textbf{r-app}
\and 
\rstlcsub
  \end{mathpar}

\fbox{  $\jsubtype{\tau}{\tau'}$ } 

\begin{mathpar}
    \inferrule{
    }
    {
      \jsubtype{\trbool}{\tubool}
    }~\textbf{bool}
    \and
     \inferrule{
       \jsubtype{\tau_1'}{\tau_1} \\
       \jsubtype{\tau_2}{\tau_2'}
     }
     {
       \jsubtype{\tau_1 \to \tau_2}{\tau_1' \to \tau_2'}
     }~\textbf{$\to$}
     \and
     \inferrule
     {
     }
     {  \jsubtype{\tau}{\tau}
     }\textbf{refl}
     \and
     \inferrule{
       \jsubtype{\tau_1}{\tau_2} \\
       \jsubtype{\tau_2}{\tau_3}
     }
     {
       \jsubtype{\tau_1}{\tau_3}
     }\textbf{trans}
   \end{mathpar}

 \caption{\relstlc typing and subtyping rules}
\label{fig:rel-stlc-typing-subtyping}
\end{figure}




\paragraph{Algorithmic (bidirectional) typing}
The type system presented above is declarative, \ie\ it doesn't
prescribe an algorithm for building a typing derivation. In fact, two
aspects of {\relstlc} make it difficult to straightforwardly
algorithmize. First, {\relstlc} doesn't have explicit type annotations
on variable bindings, which makes the system non syntax-directed:
Reading the rule \textbf{r-app} bottom-up, the argument type $\tau_1$
must be guessed. Second, the \textbf{trans} subtyping rule is also not
syntax-directed (the type $\tau_2$ must be guessed). Hence, the typing
and subtyping rules of {\relstlc} cannot be directly interpreted as a
typechecking algorithm.

A well-established way of making typing rules syntax-directed (hence
obtaining a typechecking algorithm) is to make the rules
\emph{bidirectional}~\cite{Pierce:2000,xi99:dependent,XiThesis}. In
comparison to fully annotating all binders, which could be tedious
for a programmer, the main idea behind bidirectional typechecking is
to only annotate programs at the top-level and at explicit
$\beta$-redexes (which are usually rare) and \emph{infer} all other
types.

In the case of {\relstlc}, bidirectional typechecking splits the usual
typing judgment $\Gamma \jstype{e_1}{e_2}{\tau}$ into two judgments:
(1) the checking judgment $\Gamma \vdash e_1 \backsim e_2 \downarrow
\tau$, where the type $\tau$ is an input (the type is checked), and
(2) the inference judgment $\Gamma \vdash e_1 \backsim e_2 \uparrow
    {\color{red}\tau}$, where the type $\tau$ is an output (the type
is inferred). As a convention, we write all outputs in
\textcolor{red}{red} and all inputs in
black. \Cref{fig:rel-stlc-alg-typing-subtyping} shows selected
algorithmic typing rules. We explain below the basic principles behind
the bidirectional typing rules. These principles are completely
standard for \emph{unary} type systems~\cite{xi99:dependent,XiThesis};
our observation thus far is simply that they apply as-is to relational
type systems as well and, for {\relstlc}, they suffice to ensure
completeness of bidirectional typechecking (this will cease to be the
case for later type systems).

\noindent- Types of variables and elimination forms are inferred
  (e.g., rules 
  \textbf{alg-r-app}, \textbf{alg-r-if}) whereas types of
  introduction forms are checked (e.g., rule
  \textbf{alg-r-lam}).
  
\noindent- In checking mode, the rule \textbf{alg-r-$\uparrow\downarrow$}
  allows switching to inference mode. The requirement is that the
  inferred type must be a subtype of the checked type.
  
\noindent- In inference mode, it is permissible to switch to checking mode
  when an expression's type has been explicitly annotated by the
  programmer (rule \textbf{alg-r-anno-$\uparrow$}). It can be shown
  that, for completeness, it suffices to annotate only at explicit
  $\beta$-redexes (although there is no prohibition on annotating at
  other places).

Subtyping also has an algorithmic counterpart,
$\jalgssubtype{\tau_1}{\tau_2}$, shown in
\Cref{fig:rel-stlc-alg-typing-subtyping}.  We introduce two additional
rules for reflexivity of base types (rules \textbf{alg-bl-u} and
\textbf{alg-bl-r}). Importantly, it can be proved that reflexivity and
transitivity of subtyping are \emph{admissible}, so, in particular,
there is no need for an explicit rule of transitivity, which, as
mentioned, is difficult to use in an algorithm.





\begin{figure}
 \raggedright 
\fbox{$\Gamma \infsdiff{e_1}{e_2}{\tau}$,  $\Gamma \chsdiff{ e_1}{e_2}{\tau}$ }
\begin{mathpar}
    \and
    \inferrule{\Gamma \infsdiff{e}{e'}{\trbool} \\\\
      \Gamma \chsdiff{ e_1}{e_1'}{\tau} \\
      \Gamma \chsdiff{ e_2}{e_2'}{\tau}
    }
    {\Gamma \chsdiff{\eif e \ethen e_1 \eelse e_2}{\eif e' \ethen e_1' \eelse e_2'}{\tau}}~\textbf{alg-r-if}
    \and
    \inferrule{
       \Gamma, x: \tau_1  \chsdiff{e_1}{e_2}{\tau_2}
     }
     {\Gamma \chsdiff{ \elam x.e_1}{\elam x.e_2}{
                \tau_1 \to \tau_2}}~\textbf{alg-r-lam}
    \and
    \inferrule{
      \Gamma \infsdiff{e_1}{e_1'}{\tau_1
        \to \tau_2} \\
      \Gamma \chsdiff{ e_2}{e_2'}{\tau_1}}
    {\Gamma \infsdiff{e_1 \eapp e_2}{e_1' \eapp e_2'}{\tau_2}}~\textbf{alg-r-app}
    \and
    \inferrule{
      \Gamma \infsdiff{e}{e'}{\tau'}\\
      \jalgssubtype{\tau'}{\tau} \\
    }
    { 
      \Gamma \chsdiff{e}{e'}{\tau}
    }~\textbf{alg-$\uparrow\downarrow$}
    \and
    \inferrule
    {
      \Gamma \chsdiff{e}{e'}{\tau}    
    }
    {
      \Gamma  \infsdiff{(e:\tau)}{(e':\tau)}{\tau}
    } ~\textbf{alg-r-anno-$\uparrow$} 
  \end{mathpar}

\fbox{$\jalgssubtype{\tau}{\tau'}$ } 

\begin{mathpar}
    \inferrule{
    }
    {
      \jalgssubtype{\trbool}{\trbool}
    }~\textbf{alg-bl-r}
    \and
    \inferrule{
    }
    {
      \jalgssubtype{\tubool}{\tubool}
    }~\textbf{alg-bl-u}
    \and
    \inferrule{
    }
    {
      \jalgssubtype{\trbool}{\tubool}
    }~\textbf{alg-bl}
    \and
     \inferrule{
       \jalgssubtype{\tau_1'}{\tau_1} \\\\
       \jalgssubtype{\tau_2}{\tau_2'}
     }
     {
       \jalgssubtype{\tau_1 \to \tau_2}{\tau_1' \to \tau_2'}
     }~\textbf{alg-$\to$}
   \end{mathpar}
 \caption{\relstlc algorithmic typing and subtyping rules }
\label{fig:rel-stlc-alg-typing-subtyping}
\end{figure}





The bidirectional type system's rules, when read bottom-up, can be
interpreted as a syntax-directed algorithm for typechecking. This
algorithm is sound relative to the declarative type system in the
following sense: If either $\Gamma \infsdiff{e_1}{e_2}{\tau}$ or
$\Gamma \chsdiff{ e_1}{e_2}{\tau}$, then $\Gamma
\jstype{|e_1|}{|e_2|}{\tau}$, where $|e|$ is obtained by erasing type
annotations from $e$. The bidirectional type system is also
\emph{complete} relative to the declarative type system: If $\Gamma
\jstype{e_1}{e_2}{\tau}$, then there are type-annotated variants,
$e_1'$, $e_2'$ of $e_1$, $e_2$ such that $\Gamma \infsdiff{
  e_1'}{e_2'}{\tau}$. These annotations can be limited to the
top-level and any explicit $\beta$-redexes. The proofs of these
statements are in the appendix.






\section{\relref}
\label{sec:relref}

\begin{figure*}
\raggedright
\begin{mathpar}

\inferrule
  { \ctx \jstype{e_1}{e_1'}{\tau} \\\\
    \ctx \jstype{e_2}{e_2'}{\tlist{n}{\alpha}{\tau}}
  }
  { \ctx \jstype{ \econs(e_1,e_2) }{ \econs(e_1',e_2')}{\tlist{n+1}{\alpha+1}{\tau}}}~\textbf{rr-cons1}
 \inferrule
  { \ctx \jstype{e_1}{e_1'}{\tbox{\tau}} \\\\
    \ctx \jstype{e_2}{e_2'}{\tlist{n}{\alpha}{\tau}}}
  { \ctx \jstype{ \econs(e_1,e_2) }{
      \econs(e_1',e_2')}{\tlist{n+1}{\alpha}{\tau}}}~\textbf{rr-cons2}
\and

\inferrule
  {\ctx \jstype{e}{e'}{\tlist{n}{\alpha}{\tau}} \\
    \Delta; \Phi_a \wedge n = 0  ;\Gamma \jstype{e_1}{e_1'}{\tau'} \\\\
   i, \Delta; \Phi_a \wedge n = i+1;  h: \tbox{\tau},
   tl : \tlist{i}{\alpha}{\tau}, \Gamma \jstype{e_2}{e_2'}{\tau'} ~~
\\
   i, \beta, \Delta; \Phi_a \wedge n = i+1 \wedge \alpha = \beta +1 ;
   h: \tau, tl : \tlist{i}{\beta}{\tau}, \Gamma \jstype{e_2}{e_2'}{\tau'} ~~
   }
  {\ctx \jstype{ \ecase e \eof \enil \rightarrow e_1 ~|~ h::tl \rightarrow
    e_2}{\ecase e' \eof \enil \rightarrow e_1' ~|~ h::tl \rightarrow
      e_2'}{\tau'}}~\textbf{rr-caseL}
\and
\inferrule
  {
    \Delta; \Phi_a \wedge C; \Gamma  \jstype{e_1}{e_2}{\tau}\\\\
    \Delta; \Phi_a \wedge \neg C; \Gamma  \jstype{e_1}{e_2}{\tau}\\
    \Delta \wfcs{C}
  }
  {
    \ctx  \jstype{e_1}{e_2}{\tau} } ~\textbf{rr-split}
\and  
\inferrule
{
\Delta; \Phi_a; \Gamma \jstype{e}{e}{\tau}~~\\\\
\forall x \in dom(\Gamma).~~
\Delta; \Phi_a  \jsubtype{\Gamma(x)}{\tbox{\Gamma(x)}} ~~
}
{
\Delta; \Phi_a; \Gamma,\Gamma' \jstype{e}{e}{\tbox{\tau}}
} ~\textbf{rr-nochange}
\end{mathpar}

\begin{mathpar}
     \inferrule{
     }
     {
      \Delta; \Phi_a \jsubtype{\square (\tau_1 \to \tau_2)}{ \square \tau_1 \to
         \square \tau_2}
     }\textbf{$\to\square$diff}
     \and
  \stlist
\and
\stlistzero
\and
\stlistbox
\and
\stbox
\and
%


\end{mathpar}
\caption{\relref typing and subtyping}
\label{fig:relref-ts} 
\end{figure*} 

\begin{figure*}
\raggedright


\begin{mathpar}
    \inferrule{
       {\ifextra
       \Delta; \Phi_a \wfty{\tau_1 \to \tau_2}
       \fi} \\
       \Delta; \Phi_a;  x: \tau_1, f : \tau_1 \to \tau_2, \Gamma  \jstypet{e_1}{e_2}{\tau_2}}
            {\ctx \jstypet{ \efix f(x).e_1}{\efix f(x).e_2}{
              \tau_1 \to \tau_2}}~\textbf{c-r-fix}
   \and
\inferrule
{
\Delta; \Phi_a; \tbox{\Gamma} \jstypet{e}{e}{\tau}
}
{
\Delta; \Phi_a; \tbox{\Gamma}, \Gamma' \jstypet{\enoch e}{\enoch e}{\tbox{\tau}}
} ~\textbf{c-nochange}
\and
    \inferrule
  {
    \Delta; \Phi_a \wedge C; \Gamma \jstypet{e_1}{e_2}{\tau} \quad
    \Delta; \Phi_a \wedge \neg C; \Gamma \jstypet{e_1'}{e_2'}{\tau}\\
  }
  {
    \ctx \jstypet{\esplit (e_1,e_1') \ewith C}{\esplit
      (e_2,e_2') \ewith C}{\tau} } ~\textbf{c-s}
     \inferrule
  { \ctx \jstypet{e_1}{e_1'}{\tau} \quad
    \ctx \jstypet{e_2}{e_2'}{\tlist{n}{\alpha}{\tau}}
  }
  { \ctx \jstypet{ \econsC(e_1,e_2) }{ \econsC(e_1',e_2')}{\tlist{n+1}{\alpha+1}{\tau}}}~\textbf{c-cons1}
    \and
   \inferrule
  { \ctx \jstypet{e_1}{e_1'}{\tbox{\tau}} \\
    \ctx \jstypet{e_2}{e_2'}{\tlist{n}{\alpha}{\tau}}}
  { \ctx \jstypet{ \econsNC(e_1,e_2) }{
      \econsNC(e_1',e_2')}{\tlist{n+1}{\alpha}{\tau}}}~\textbf{c-r-cons2}
\quad
    \inferrule{
     \ctx \jstypet{e}{e'}{\tau}\\
     \Delta; \Phi_a \jeqtype{\tau}{\tau'} 
  }
  {\ctx \jstypet{e}{e'}{\tau'} }
  ~{\textbf{c-r-$\equiv$}}
    
    
  \end{mathpar}
\caption{\relref core typing rules}
 \label{fig:relref-c}
\end{figure*}




\begin{figure*}
\raggedright

\begin{mathpar}
  
   \inferrule
{
\freshSize{i,\beta} \\
\Delta; {\psi_a}; \Phi_a; \Gamma  \chdifft{e_1}{e_1'}{\tau}{t_1}{\Phi_1} \\
\Delta; {i,\beta,\psi_a}; \Phi_a; \Gamma
\chdifft{e_2}{e_2'}{\tlist{i}{\beta}{\tau}}{t_2}{\Phi_2}\\
\Phi_2' = \cand{\ceq{n}{(i+1)}}{\cexists{\beta}{\ssize}{\cand{\Phi_2}{\ceq{\alpha}{\beta+1}}}}
}
{
  \bctx \chdifft{\econsC(e_1,e_2)}{\econsC(e_1',e_2')}{\tlist{n}{\alpha}{\tau}}{t}{\cand{\Phi_1}{\cexists{i}{\ssize}{\Phi_2'}}}
}~\textbf{alg-r-consC-$\downarrow$}
  \and
\inferrule
{
\freshSize{i} \\
\Delta; {\psi_a}; \Phi_a; \Gamma  \chdifft{e_1}{e_1'}{\tbox{\tau}}{t_1}{\Phi_1} \\\\
\Delta; {i,\psi_a}; \Phi_a; \Gamma
\chdifft{e_2}{e_2'}{\tlist{i}{\alpha}{\tau}}{t_2}{\Phi_2}\\
\Phi_2' = \cand{\Phi_2}{\ceq{n}{(i+1)}}
}
{
\bctx \chdifft{\econsNC(e_1,e_2)}{\econsNC(e_1',e_2')}{\tlist{n}{\alpha}{\tau}}{t}{\cand{\Phi_1}{\cexists{i}{\ssize}{\Phi_2'}}}
}~\textbf{alg-r-consNC-$\downarrow$}
  \and
    \inferrule{
  \Delta; \psi_a; C \wedge \Phi_a; \Gamma \chdifft{e_1}{e_1'}{\tau}{t}{\Phi_1}\\
  \Delta; \psi_a; \neg C \wedge \Phi_a; \Gamma \chdifft{e_2}{e_2'}{\tau}{t}{\Phi_2}\\
  \Delta  \wfcs{C}
}
{
  \bctx \chdifft{\esplit (e_1, e_2) \ewith C}{\esplit (e_1', e_2') \ewith C}{\tau}{t}{\cand{\cimpl{C}{\Phi_1}}{\cimpl{\neg C}{\Phi_2}}}
} ~\textbf{alg-r-split$\downarrow$}\and
\inferrule{
\Delta; \psi_a; \Phi_a \jalgeqtype{\tau}{\tau'}{\Phi}}
{
\Delta; \psi_a; \Phi_a \jalgeqtype{\tlist{n}{\alpha}{\tau}}{\tlist{n'}{\alpha'}{\tau'}}{\cand{\Phi}{\cand{\ceq{n}{n'}}{\ceq{\alpha}{\alpha'}}}}
}~\textbf{alg-r-list} 
\end{mathpar}
\caption{{\relrefcore} algorithmic typing and subtyping rules}
 \label{fig:relref-cea}
\end{figure*}

Next, we extend bidirectional typechecking to \emph{relational
  refinements}. Relational
refinements~\citep{Pottier03:IF,gaboardi13:dep,BGGHRS15,DuCostIt16,Cicek:2017}
express fine-grained relations between pairs of expressions. They have
been used for many different purposes ranging from information flow
control to differential privacy. We consider here a simple setting,
which still suffices to bring out key challenges in applying
bidirectional typechecking to relational refinements.

We extend {\relstlc} with primitive lists and a relational refinement
type $\tlist{n}{\alpha}{\tau}$, which ascribes a pair of lists, both
of length $n$, that differ pointwise in at most $\alpha$ positions
($n$ and $\alpha$ are natural numbers). $\tlist{n}{\alpha}{\tau}$
\emph{refines} the standard list type with $n$ and $\alpha$ and the
refinement is \emph{relational} since $\alpha$ expresses a constraint
on the two lists together. To construct lists of this type when
$\alpha \not= n$, we also need a way to express that at least
$(n-\alpha)$ elements are pointwise equal. To this end, we introduce
the comonadic type $\tbox{\tau}$, which ascribes pairs of expressions
of type $\tau$ that are equal (\ie\ the diagonal relation on
$\tau$). Type-level terms like $n$ and $\alpha$ are called index terms
or indices, generically denoted $I$. The type system also supports
quantification over such terms. To write recursive programs on lists,
we also add a fixpoint operator, which poses no additional
difficulty for bidirectional typechecking.
The resulting system, called {\relref}, has the following syntax.
\[
\begin{array}{@{}l@{\,}l@{}}
 \label{fig:syntax-relref}
   \mbox{Types} 
  \    \tau ::= & \dots ~|~ \tlist{n}{\alpha}{\tau} ~|~ \tsforall{i}
                  \tau ~|~ \texists{i} \tau  \\ 
  &~|~ \tbox{\tau}  ~|~ \tcprod{C}{\tau}
                  ~|~ \tcimpl{C}{\tau} \\

  \mbox{Expr}  \ e ::= & \dots ~|~ \efix
                                 f(x).e  ~|~ \enil
                                 ~|~ \econs(e_1,e_2) \\ 
& ~|~ (\!\ecase e \eof \enil \rightarrow e_1 
 |~ h ::tl \rightarrow e_2)
~|~ \eLam e  \\ & ~|~e \eApp ~|~ \elet x = e_1 \ein e_2 ~|~  \clet e_1 \eas x
    \ein e_2 \\ 
&~|~  \epack e
                              ~|~  \eunpack e_1 \eas x \ein e_2 ~|~ \ecelim e  \\
\mbox{Indices}\ 
 I, n,\alpha ::= &  
i \bnfalt
\szero \bnfalt 
\splusone{I} \bnfalt
\splus{I_1}{I_2} \bnfalt
\sminus{I_1}{I_2} \bnfalt 
\sdiv{I_1}{I_2} \bnfalt \\
& \smult{I_1}{I_2} \bnfalt
\sceil{I} \bnfalt
\sfloor{I} \bnfalt
\smin{I_1}{I_2} \bnfalt
\smax{I_1}{I_2}
\end{array}
\] 


Types can quantify over index variables, $i$, as in $\tsforall{i}
\tau$ and $\texists{i} \tau$. The constructs $\epack e$ and $\eunpack
e_1 \eas x \ein e_2$ are the introduction and elimination forms for
existentially quantified types. The constructs $\eLam. e$ and $e
\eApp$ are the introduction and elimination forms for universally
quantified types. To represent arithmetic relations over index
variables, constraints denoted $C$, sets of predicates over index
terms, appear in types as in $\tcprod{C}{\tau}$ and
$\tcimpl{C}{\tau}$. The type $\tcprod{C}{\tau}$ means the type $\tau$
and that $C$ holds, while $\tcimpl{C}{\tau}$ means that, if the
constraint $C$ holds, then the type is $\tau$. The construct $\clet
e_1 \eas x \ein e_2$ is the elimination form for the constrained type
$\tcprod{C}{\tau}$. By design, index terms do not appear in {\relref}
expressions.

\paragraph{Example (map)}
As an example, we can write the standard list map function, and give it a very informative relational type in {\relref}.
\begin{nstabbing}
 $\efix \kw{map}(f). \eLam. \eLam. \elam l. \ecase l $\=$\eof \enil \rightarrow \enil$ \\
\>$ |~ h :: tl \rightarrow \econs(f \eapp h, \eapp \kw{map} \eapp f \eApp \eApp \eapp tl) $
\end{nstabbing}
\medskip
\begin{equation*}
\kw{map}: (\tbox{(\tau_1 \to \tau_2)}) \to \forall
  n,\alpha. \tlist{n}{\alpha}{\tau_1} \to
  \tlist{n}{\alpha}{\tau_2}\label{eq:map}
\end{equation*}
The type means that two runs of map with \emph{equal} mapping
functions and two lists that differ in at most $\alpha$ positions
result in two lists with the same property. Notice how $\alpha$ is
universally quantified in the type, and how $\tbox{}$ represents that
the mapping function be equal in the two runs.

\paragraph{Declarative typing} {\relref}'s typing judgment has the form
$\Delta; \Phi_a; \Gamma \jstype{e_1}{e_2}{\tau}$ and means that $e_1$
and $e_2$ have the relational type $\tau$ if the constraints $\Phi_a$
hold. $\Delta$ is a (universally quantified) context of index
variables and $\Gamma$, as usual, is the typing context for program
variables. \Cref{fig:relref-ts} shows selected typing rules that use
refinements and constraints in interesting ways, and make
bidirectional typechecking difficult.
There are \emph{two} rules for typing the list $\econs$
constructor. Rule \textbf{rr-cons1} applies when the head elements of
the constructed lists may differ. Note how the relational refinement
$\alpha$ changes to $\alpha+1$ from the premise to the
conclusion. Rule \textbf{rr-cons2} applies when the head elements are
equal, witnessed by the comonadic type $\tbox{\tau}$. $\alpha$ does
not change in this rule. Dually, the $\econs$ branch of list case
analysis (rule \textbf{rr-caseL}) is typed \emph{twice} with different
index constraints---once for each of these two possible ways of
constructing the $\econs$-ed list. A consequence of this double typing
of the same branch with different constraints is that expressions
cannot contain index terms (else such typing may be impossible). The
rule \textbf{rr-split} case-splits on an arbitrary constraint $C$ in
the context. This is useful for typing recursive
functions~\cite{DuCostIt16,Cicek:2017}. Finally, the rule for
introducing the type $\tbox{\tau}$, \textbf{rr-nochange}, is
interesting. It says that if $e$ relates to itself at type $\tau$ and
all variables in $\Gamma$ morally have $\tbox{}$-ed types (checked via
subtyping), then $e$ also relates to itself at type $\tbox{\tau}$.

\paragraph{Declarative subtyping}
{\relref} subtyping $\tau \sqsubseteq \tau'$ is complex. Some of the
rules are shown in Figure~\ref{fig:relref-ts}.  First, subtyping is
constraint-dependent, because it must, for instance, be able to show
that $\tlist{n}{\alpha}{\tau} \sqsubseteq \tlist{m}{\alpha}{\tau}$
when $n = m$. Second, in {\relref}, $\tbox{}$'s comonadic properties
manifest themselves via subtyping. This results in interactions
between $\tbox{}$ and other connectives as, for instance, in the rules
\textbf{$\to \square \wdiff$}, \textbf{l2} and \textbf{l$\tbox{}$}.

We explain some of the subtyping rules. The rule \textbf{l1} allows
the number of elements that differ in two lists to be weakened
(covariantly). The rule \textbf{l2} allows two related lists with zero
differences to be retyped as two related lists whose elements are in
the diagonal relation. The rule \textbf{l$\square$}
allows two related lists whose elements are equal to be
retyped as two equal lists, represented by the outer $\square$.  The
rule \textbf{T} coerces $\tbox{\tau}$ to $\tau$ by forgetting that the
two related elements are, in fact, equal.


\paragraph{Towards algorithmization}
An algorithm for typechecking {\relref} faces two difficulties beyond
those seen in {\relstlc}. Both difficulties arise due to {\relref}'s
relational refinements. First, there is additional
non-syntax-directedness in the rules: Rules \textbf{rr-cons1} and
\textbf{rr-cons2} apply to expressions of the \emph{same} shape (the
rules differ in their treatment of index terms), and rules
\textbf{rr-split} and \textbf{rr-nochange} are not syntax-directed
(their use overlaps with other rules). Second, owing to the
interaction between $\tbox{}$ and other type constructs, it is
infeasible to re-define subtyping in a way that makes transitivity
admissible. As a result of these two problems, bidirectional typing
alone does not yield an \emph{algorithm} for typechecking.

An obvious way to address the first of these problems is to force
additional annotations in expressions to remove the
non-syntax-directness. However, this will not address the problem with
subtyping. Importantly, it also will not allow us to formally connect
the algorithmic type system to the declarative type system above (with
non-syntax-directedness).

Consequently, we follow a slightly different approach here. First, we
introduce a simpler core calculus {\relrefcore}, which annotates
expressions to resolve the lack of syntax-directedness in typing
rules. Additionally, {\relrefcore} features only \emph{type
  equivalence}, not subtyping. We show that every {\relref} expression
can be \emph{elaborated} to a semantically equivalent expression in
{\relrefcore} by adding enough annotations and expressing subtyping as
definable type coercions. Next, we build a bidirectional, algorithmic
type system for {\relrefcore} and prove it relatively sound and
complete. End-to-end, this makes a strong theoretical point: There is
a calculus (\relrefcore) that is as expressive as {\relref}, and that
is fully amenable to bidirectional typechecking.


\paragraph{\relrefcore syntax}
The new calculus {\relrefcore} is similar to {\relref} but has
explicit syntactic markers to indicate which typing rules to apply
where, thus resolving the nondeterminism caused by the aforementioned
typing rules such as \textbf{rr-split} and \textbf{rr-cons1/rr-cons2}.
The expression syntax of {\relrefcore} is as follows.
\[
\begin{array}{ll}
 \label{fig:syntax}
  \mbox{Expr} \  e  ::=  & \dots 
    ~|~ \esplit(e_1,e_2) \ewith C ~|~ \enoch e
  ~|~ \eLam i. e
                                 ~|~ e [I] ~|\\  
  & \econsNC(e_1,e_2) ~|~ \econsC(e_1,e_2) ~|
  \\

 &                      \left(\begin{minipage}[c]{13em}
                           $\ecase e \eof \enil \, \rightarrow e_1$ \\
                           $|~h\,::_{NC} tl\, \rightarrow e_2
                           ~|~h::_{C} tl~ \rightarrow e_3$
            \end{minipage}\right)  
\end{array}
\]   
The list constructor $\econs$ is separated into two---$\econsC$ and
$\econsNC$---to disambiguate the rules \textbf{rr-cons1} and
\textbf{rr-cons2}.  Correspondingly, the list-case construct now has
three branches---one each for $\enil$, $\econsC$ and
$\econsNC$. ``$\esplit(e_1,e_2) \ewith C$'' indicates that rule
\textbf{rr-split} must be applied and records the constraint $C$ to
split on. To write meaningful $C$s, expressions now carry index
terms. For instance, the elimination form for universally quantified
types in {\relrefcore} is $e [I]$ as opposed to {\relref}'s $e
\eApp$. The construct $\enoch e$ indicates an application of the rule
\textbf{rr-nochange}.

\paragraph{\relrefcore typing rules}
Selected rules of {\relrefcore}'s typing judgment
$\Delta;\Phi_a;\Gamma \jstypet{e}{e'}{\tau}$ are shown in
\Cref{fig:relref-c}. $\Phi_a$ and $\Delta$ are interpreted as in
\relref. {\relrefcore}'s rules are similar to {\relref}'s, but there
are important differences. First, rules are now
syntax-directed. Second, there is no relational subtyping. Instead,
there is type-equivalence, $\pmb{\equiv}$, which is much simpler than
subtyping. It only lifts equality modulo constraints to types (e.g.,
$\tlist{1+2}{\alpha}{\tau} \pmb{\equiv} \tlist{3}{\alpha}{\tau}$) and
it can be easily implemented algorithmically (modulo constraint
solving). We omit its straightforward details.

\paragraph{Simulating {\relref}'s subtyping}
A key property of {\relrefcore} is that it can simulate {\relref}'s
subtyping via explicit coercion functions, as formalized in the
following lemma. Such elimination of subtyping is a common technique
for simplifying typechecking in the unary
setting~\cite{Breazu-Tannen:1991,Crary:2000}; here, we lift the idea
to the relational setting and to our comonad.


\smallskip
\begin{lem}\label{lem:relref-elab}
  If $\Delta; \Phi_a \jsubtype{\tau}{\tau'}$ in {\relref} then there exists
  $e \in \mbox{\relrefcore}$ s.t.\ $\Delta; \Phi_a;
  \cdot
  \jstypet{e}{e}{\tau
    \to \tau'}$.
\end{lem}
\begin{proof}
 By induction on the subtyping derivation.
\end{proof}

\paragraph{Elaboration}
Given \Cref{lem:relref-elab}, we define a straightforward
type derivation-directed embedding from {\relref} to
{\relrefcore}. Briefly, we use the {\relref} typing derivation to
insert additional syntactic annotations that {\relrefcore} needs and
use \Cref{lem:relref-elab} wherever subtyping appears in the {\relref}
derivation. The embedding preserves well-typedness (see the appendix
for details). This shows that {\relrefcore} is as expressive as
{\relref}. (In fact, this is expressiveness in a strong sense, dubbed
macro-expressiveness by
Felleisen~\cite{DBLP:journals/scp/Felleisen91}.)

\paragraph{Algorithmic (bidirectional) typechecking}
We now build an algorithmic, bidirectional type system for
{\relrefcore}. We call this system {\birelref}.
%
Selected rules of {\birelref} are shown in \Cref{fig:relref-cea}.  As
before, {\birelref} has two typing judgments: one to check types and
the other to infer them. The key addition over {\relstlc} is that
{\birelref}'s typing judgments output \emph{constraints} between index
terms, which must be verified for typing.  The checking judgment has
the form $\Delta; \psi_a; \Phi_a; \Gamma
\chdifft{e_1}{e_2}{\tau}{t}{\Phi}$. Here, the type $\tau$ is an input,
but the constraints $\Phi$ are an output. The intuitive meaning is
that \emph{if constraints \textcolor{red}{$\Phi$} hold} (assuming
$\Phi_a$), then $\Delta;\Phi_a;\Gamma \jstypet{e_1}{e_2}{\tau}$ holds
in {\relrefcore}. (The reader may ignore the index variable context
$\psi_a$; it contains variables universally quantified in $\Phi$.)  As
an example, consider the rule \textbf{alg-r-consC-}$\downarrow$ for
relating cons-ed lists at type $\tlist{n}{\alpha}{\tau}$ when the
heads may differ. To do this, the rule relates the tails at type
$\tlist{i}{\beta}{\tau}$ for some $i$ and $\beta$. For this to be
sound, $\ceq{n}{i+1}$ and $\ceq{\alpha}{\beta+1}$ must hold, so these
appear as constraints in \textcolor{red}{$\Phi$}. Note that
\textcolor{red}{$\Phi$} quantifies over $i$ and $\beta$
existentially. Constraint solvers cannot handle such existential
quantification easily, a point to which we return in
Section~\ref{sec:implementation}.

In the inference judgment $\Delta; \psi_a; \Phi_a; \Gamma
\infdifft{e_1}{e_2}{\tau}{\psi}{t}{\Phi}$, both $\Phi$ and $\tau$ are
outputs. The meaning of the judgment is similar: if
\textcolor{red}{$\Phi$} holds (assuming $\Phi_a$), then
$\Delta;\Phi_a;\Gamma \jstypet{e_1}{e_2}{\tau}$.

Like typing, the algorithmic type equivalence judgment $\Delta;
\psi_a; \Phi_a \jalgeqtype{\tau}{\tau'}{\Phi}$ also generates
constraints.

\paragraph{Soundness and completeness}
We prove that {\birelref} is sound and complete
w.r.t.\ {\relrefcore}'s declarative type system. Soundness says that
any inference or checking judgment provable in the algorithmic type
system can be simulated in {\relrefcore} if the output constraints
\textcolor{red}{$\Phi$} are satisfied. Dually, completeness says that
any pair of typeable {\relrefcore} programs can be sufficiently
annotated with types to make their type checkable in {\birelref},
with satisfiable output constraints. As before, we write $\erty{e}$
for the erasure of typing annotations from a {\birelref} expression
$e$ to yield a {\relrefcore} expression.

\smallskip
\begin{thm}[Soundness]\label{thm:alg-typing-soundness-rr}
\mbox{}
  \begin{enumerate}
\item Assume $\bctx \chdifft{e}{e'}{\tau}{t}{\Phi}$,
  $\fivars{\Phi_a,\Gamma,\tau} \subseteq dom(\Delta, \psi_a)$ ,
  $\theta_a$ is a valid substitution for $\psi_a$ s.t.
  $\Delta; \suba{\Phi_a} \sat \suba{\Phi}$ holds. Then,
  $\Delta; \suba{\Phi_a}; \suba{\Gamma}
  \jtypediffMt{\suba{t}}{\erty{e}}{\erty{e'}}{\suba{\tau}}$.

\item Assume $\bctx \infdifft{e}{e'}{\tau}{\psi}{t}{\Phi}$,
  $\fivars{\Phi_a,\Gamma} \subseteq dom(\Delta, \psi_a) $,$
  \theta_a$ is a valid substitution for $\psi_a$
  s.t. $\Delta; \suba{\Phi_a} \sat \suba{\Phi}$ holds. Then,
  $\Delta; \suba{\Phi_a}; \suba{\Gamma}
  \jtypediffMt{\subta{t}}{\erty{e}}{\erty{e'}}{\suba{\tau}}$.
\end{enumerate}  
\end{thm}
\begin{proof}
By simultaneous induction on the given {\birelref} derivations.
\end{proof}

\begin{thm}[Completeness]\label{thm:relref-alg-typing-completeness}
\mbox{}
  \begin{enumerate}
  \item Assume that $\Delta; \Phi_a; \Gamma
    \jtypediffMt{t}{e_1}{e_2}{\tau}$.  Then, there exist $e_1',e_2'$
    such that $\Delta; \cdot; \Phi_a; \Gamma
    \chdifft{e_1'}{e_2'}{\tau}{t}{\Phi}$ and $\Delta; \Phi_a \sat \Phi$
    and $\erty{e_1'} = e_1$ and $\erty{e_2'} = e_2$.
  \end{enumerate}  
\end{thm}
\begin{proof}
By induction on the given {\relrefcore} typing derivation.
\end{proof}

\section{\relinfref}
\label{sec:relinfref}

In {\relref}, all rules analyze the two (related) expressions
simultaneously. This limits the type system to pairs of expressions
that are structurally similar. In many cases, however, it is possible
to prove a relation between two dissimilar expressions by analyzing
them \emph{individually}. For example, if we can prove that $e_1$ and
$e_2$ individually produce lists of length $n$ with elements of type $\tau$,
then it is immediate that
$\jstype{e_1}{e_2}{\tlist{n}{n}{\tau}}$. We now extend {\relref} to
allow falling back to such \emph{unary} analysis. We call the new
system {\relinfref}.

{\relinfref} adds a new class of unary types, $\grt$, which ascribe
individual expressions. These are the ``standard'' types from existing
refinement systems like DML~\cite{xi99:dependent}. For example, the
unary type for lists, $\ulist{n}{\grt}$, carries a refinement $n$, the
length of the list. Unary types also include quantification over index
variables, which we elide here for brevity. Importantly, we also add a
new relational type $\tch{\grt_1}{\grt_2}$ which ascribes any pair
$e_1$, $e_2$, whose unary types are $\grt_1$ and $\grt_2$,
respectively.
%
\[
\begin{array}{llll}
 \label{fig:syntax_relinfref}
  \mbox{Unary types} 
  & \grt & ::= & \tbool ~|~ \grt_1 \to \grt_2 ~|~ \ulist{n}{\grt} ~|~ \ldots \\
  \mbox{Relational types} 
  &  \tau & ::= & \dots ~|~ \tch{\grt_1}{\grt_2} 
 \end{array}
\]

\paragraph{Declarative typing}
{\relinfref} has two typing judgments, unary and relational. The unary
judgment's rules are exactly those of standard (unary) refinement type
systems like DML~\cite{xi99:dependent}, so we elide them here. The
relational rules are those of {\relref} and the following new rule,
\textbf{r-switch}, which allows the use of unary typing in relational
typing. Here, $|.|_i$ is a projection function that converts a
relational type to its left ($i=1$) or right ($i=2$) unary type by
forgetting relational refinements. For example, $|\tch{\grt_1}{\grt_2}|_1 = \grt_1$ and $|\tlist{n}{\alpha}{\tau}|_1 =
\ulist{n}{(|\tau|_1)}$.
%
\[ 
\inferrule
    {  |\Gamma|_1 \jtype{}{}{e_1}{\grt_1}   \\ |\Gamma|_2 \jtype{}{}{e_2}{\grt_2}  }  
    { \Gamma \jstype{e_1}{e_2}{  \tch{\grt_1}{\grt_2} }}~\textbf{r-switch}
\]

Besides this rule, the second interesting aspect of {\relinfref} is
subtyping for $\tch{\grt_1}{\grt_2}$, which makes unary typing useful
in relational reasoning. For instance, the example given at the
beginning of this section is typed using \textbf{r-switch} and the
subtyping rule $\tch{\ulist{n}{\grt_1}}{\ulist{n}{\grt_2}} \sqsubseteq
\tlist{n}{n}{(\tch{\grt_1}{\grt_2})}$.

\paragraph{Algorithmization}
Algorithmizing {\relinfref} faces new hurdles: The new rule
\textbf{r-switch} is also not syntax-directed and subtyping for
$\tch{\grt_1}{\grt_2}$ is complex and cannot be re-defined to make
transitivity admissible. Consequently, we follow the approach we used
for {\relinf}. We define a core language, {\relinfrefcore}, that has
the syntactic markers of {\relrefcore} and a new construct $\eswitch
e$ (which marks the rule \textbf{r-switch}) to resolve the ambiguity
in typing rules. The language has simple type equivalence, not
subtyping. We then define an elaboration of {\relinfref} into
{\relinfrefcore}. Finally, we define a bidirectional, algorithmic type
system called {\birelinfref} and prove it sound and complete relative
to {\relinfrefcore}.  The entire development is not particularly more
challenging than that of {\relinf} but is more tedious because we have
to handle both unary and relational typing. Conceptually, the
interesting aspect is that, in typing examples, we found it convenient
to apply the \textbf{r-switch} rule in both checking and inference
mode in the bidirectional type system, so this type system features
two versions of the rule, one in each mode. This does not cause
ambiguity in the rules since the mode is always uniquely known.

\section{\tname}
\label{sec:relcost}

As our last step, we add a relational effect, namely, relative cost to
{\relinfref}. This results in the type system {\tname} of
\citet{Cicek:2017}. {\tname} allows establishing an upper bound on the
\emph{relative cost} of two expressions $e_1$ and $e_2$, \ie\ on
$\mbox{cost}(e_1) - \mbox{cost}(e_2)$. For this, {\tname} extends
     {\relinfref}'s types and judgments with cost effects. Expressions
     are syntactically those of {\relinfref}, but the evaluation of
     elimination forms like list-case and function application
     produces non-trivial cost in the operational semantics.

We start with the types. The relational function type $\tau_1 \to
\tau_2$ is refined to $\tau_1 \tarrd{t} \tau_2$, where $t$ (an
index term of type real) is an upper-bound on the relative costs of
the bodies of the two functions that the type ascribes. Similarly, the
\emph{unary} function type $\grt_1 \to \grt_2$ is refined to $\grt_1
\uarr{k}{t} \grt_2$, where $k$ and $t$ are lower and upper bounds on
the cost of the body of the function. Other than this, the types match
those of {\relinfref}.

As an example, the list $\kw{map}$ function from
Section~\ref{sec:relref} can be given the following more informative
type in {\tname}:
\begin{equation*}
\forall t.(\tbox{(\tau_1 \tarrd{t} \tau_2)}) \to \forall
  n,\alpha. \tlist{n}{\alpha}{\tau_1} \tarrd{t \cdot \alpha}
  \tlist{n}{\alpha}{\tau_2}\label{eq:1}
\end{equation*}
This type says that if two runs of $\kw{map}$ are given the
\emph{same} mapping function whose body's cost may vary by at most $t$
across inputs, and two lists that differ in no more than $\alpha$
elements, then the relative cost of the two runs is no more than
$t\cdot\alpha$. Intuitively, this makes sense. The two runs differ
only in applications of the mapping function to list elements that
differ. Each such application results in a relative cost at most $t$
and there are at most $\alpha$ such applications.

 \begin{figure*}
    \centering
    \begin{mathpar}
      \tyrfix
      \and
      \tyrappD
      \and
        \tynochange
        \and
      \tyswitch
    \end{mathpar} 
    \caption{{\tname} relational typing rules}
    \label{fig:synch-rel-typing-rules}
  \end{figure*}

\begin{figure*}
\begin{mathpar}
  \chRnochange
  \and
 \chRupdown
\and
 \infRapp
\end{mathpar}

\caption{{\bitname} algorithmic typing rules}
\label{fig:selected-alg-typing-rules} 
\end{figure*} 

\paragraph{Declarative typing}
Like {\relinfref}, {\tname} has two typing judgments, one unary and
one relational. The difference from {\relinfref} is that these
judgments now carry \emph{cost effects}---upper and lower bounds on
the cost of the expression being typed in the unary judgment and an
upper bound on the relative cost of the two expressions in the
relational judgment. The unary judgment, $\octx
\jtype{k}{t}{e}{\grt}$, means that $e$ has the unary type $\grt$ and
its cost is upper- and lower-bounded by $t$ and $k$, respectively. The
relational judgment, $\ctx \jtypediff{t}{e_1}{e_2}{\tau}$, means that
$e_1$, $e_2$ have the relational type $\tau$ and their relative cost
is upper-bounded by $t$. Here, we describe only on the relational
judgment, since the unary judgment has simpler rules.

The rules of the relational judgment are obtained by augmenting the
rules of {\relinfref} to track costs. Selected, interesting rules are
shown in \Cref{fig:synch-rel-typing-rules}. The rule \textbf{r-fix}
types two recursive functions at type $\tau_1 \tarrd{t} \tau_2$ if
their bodies have the relative cost $t$. Rule \textbf{r-app} relates
$e_1 \eapp e_2$ and $e_1' \eapp e_2'$ with a relative cost obtained by
\emph{adding} the relative costs of $(e_1, e_1')$, $(e_2, e_2')$ and
of the bodies of the applied functions (obtained from the function
type in the first premise). Rule \textbf{nochange} says that if $e$
relates to itself in a context that has only variables of $\tbox$-ed
types (\ie\ they will get substituted by equal values in the two
runs), then $e$'s cost relative to itself is $0$. Finally, rule
\textbf{switch} allows a fallback to unary reasoning: If $e_1$'s unary
cost is upper-bounded by $t_1$ and $e_2$'s unary cost is lower-bounded
by $k_2$, then $e_1$, $e_2$'s relative cost is (trivially)
upper-bounded by $t_1-k_2$.

\paragraph{Declarative subtyping}
{\tname}'s subtyping is directly based on {\relref} and {\relinfref},
but the rules are additionally aware of costs. For example, the
{\relref} rule \textbf{$\to\square$diff} (\Cref{fig:relref-ts}) gets
refined to: $\tbox{(\tau_1 \tarrd{t} \tau_2)} \sqsubseteq
\tbox{\tau_1} \tarrd{0} \tbox{\tau_2}$, which intuitively means that
two equal functions when given two equal arguments reduce with exactly
the same cost. 

\paragraph{Towards algorithmization}
{\tname} inherits all the non syntax-directedness and subtyping
complexity of {\relinfref}, and additionally adds costs. To build an
algorithmic type system for {\tname}, we follow the approach of
({\relref} and) {\relinfref}. We first define a simpler core language,
{\tnamemin}, which resolves all rule ambiguity and has type
equivalence in place of subtyping, and elaborate {\tname} into this
core language. This step is not significantly harder than for
{\relinfref} since {\tname} does not add more rule-ambiguity. Hence,
we do not describe this step further.

The interesting step is the second one---the bidirectional type system
for {\tnamemin}. This bidirectional system uses constraints to relate
not just type refinements but also costs of subexpressions, as we
explain next.

\paragraph{Algorithmic (bidirectional) typechecking}
As before, the bidirectional type system, called {\bitname}, uses two
judgments---one for type checking and one for type inference. The
checking judgment $\Delta; \psi_a; \Phi_a; \Gamma
\chdiff{e_1}{e_2}{\tau}{t}{\Phi}$ means that if
\textcolor{red}{$\Phi$} holds (assuming $\Phi_a$), then $\ctx
\jtypediff{t}{e_1}{e_2}{\tau}$ is provable in {\tnamemin}. Note that,
like the type $\tau$, the relative cost $t$ is also an input in this
judgment, \ie\ it is also checked. The inference judgment $\Delta;
\psi_a; \Phi_a; \Gamma \infdiff{e_1}{e_2}{\tau}{\psi}{t}{\Phi}$ has a
similar meaning, but now the cost $t$ (like the type $\tau$) is an
output in the judgment, \ie\ it is also inferred. This judgment also
has an additional output $\psi$. This is a set of existentially
quantified cost variables generated by the rules. Substitutions for
these variables must be found to satisfy~$\Phi$.

To understand the need for $\psi$, let us examine a few typing rules
(\Cref{fig:selected-alg-typing-rules}). Consider the rule
\textbf{alg-r-app-$\uparrow$} for function application. In this rule,
the relative cost \textcolor{red}{$t_1$} of the functions is inferred
(first premise), but the cost of the arguments must be checked. Since
this cost is not available upfront, the rule creates a fresh
existential variable $t_2$ and ``checks'' the arguments against
that. This results in appropriate constraints on $t_2$ getting added
to \textcolor{red}{$\Phi_2$} in the second premise. $t_2$ is included
in the total cost and, importantly, it is added to
\textcolor{red}{$\psi$} in the conclusion to indicate that it is
existentially quantified (any substitution for it that satisfies the
final constraints is okay). This pattern of creating existential
variables for the costs that need to be checked but aren't already
known is pervasive in the rules, and is the key increment in
{\bitname} relative to {\birelinfref}.

We comment on two other interesting rules. The rule
\textbf{alg-r-$\uparrow\downarrow$}, which switches from inference to
checking mode when read top-down, closes the existential variables
\textcolor{red}{$\psi$} in the premise by explicitly introducing an
existential quantifier over them in the conclusion. In the rule
\textbf{alg-r-nochange-$\downarrow$} (which implements the rule
\textbf{nochange} from \Cref{fig:synch-rel-typing-rules}), the final
cost must be $0$. So, a constraint equating the given cost $t$ to $0$
is generated.

{\bitname} is sound and complete relative to {\tnamemin} in the sense
of Theorems~\ref{thm:alg-typing-soundness-rr}
and~\ref{thm:relref-alg-typing-completeness} (appropriately adapted to
the judgments of {\tname}). We defer the details to the appendix.

\paragraph{Summary}
Since bidirectional typechecking for effects has received relatively
little attention even in the context of unary analysis, we briefly
recapitulate the insights we gained from designing {\bitname}. First,
bidirectional typechecking extends very well to type systems with
effects, even when combined with refinements and relational
reasoning. Second, the mode of the effect (cost in our case) seems to
mirror the mode of the type: The checking judgment checks both the
type and the cost, while the inference judgment infers both. We did
not find the need for a judgment that checks one but infers the
other. Finally, bidirectional typechecking generates more existential
variables than it would without effects, but effects do not complicate
the meta-theory (soundness and completeness) substantially.

\section{Implementation}
\label{sec:implementation}

We have implemented a bidirectional typechecker for {\tname} in
OCaml. The typechecker implements the checking and inference judgments
of {\bitname} but, to avoid overburdening the programmer, it works on
the compact terms (expressions) of {\tname} rather than the elaborate,
annotated terms of {\tnamemin}. Hence, the terms do not resolve all
the ambiguities of which (sub)typing rules to apply when. For this,
the typechecker uses heuristics that we designed carefully by looking
at a variety of examples. Conceptually, these heuristics are a sound
but incomplete implementation of the elaboration (embedding) from
{\tname} to {\tnamemin}.

The constraints output by the bidirectional rules are solved using a
combination of a custom procedure to eliminate existential variables
and an off-the-shelf SMT solver. We describe both the heuristics and
the constraint solving below.

Note that {\tname} is a conservative extension of {\relref} and
{\relinfref}. Any derivation in {\relref} or {\relinfref} can be
simulated in {\tname} by adding the trivial cost upper-bound of
$\infty$. As a result, our implementation can also be used for
{\relref} and {\relinfref}.

\paragraph{Heuristics}
We list below the main heuristics we use to reduce the nondeterminism
in picking (sub)typing rules.
%
%
\begin{enumerate}[leftmargin=0cm,itemindent=.5cm,labelwidth=\itemindent,labelsep=0cm,align=left,itemsep=7pt,topsep=7pt]
\item \label{heur:cons} When typing a pair of $\econs$-ed lists, we
  apply the bidirectional analogues of both the rules \textbf{rr-cons1}
  and \textbf{rr-cons2} (\Cref{fig:relref-ts}) and combine the
  resulting constraints via disjunction.

\item\label{heur:split} When typing a function that takes an argument
  of type $\tlist{n}{\alpha}{\tau}$, we immediately apply the
  algorithmic analogue of the rule \textbf{rr-split}
  (\Cref{fig:relref-ts}) with $C = (\ceq{\alpha}{0})$. For the case
  $\ceq{\alpha}{0}$, we first try to complete the typing by invoking
  the rule \textbf{alg-r-nochange-$\downarrow$}
  (\Cref{fig:selected-alg-typing-rules}). This is because we found
  experimentally that many recursive list programs require this
  analysis. Moreover, the rule \textbf{rr-split} is \emph{invertible}:
  Applying the rule cannot cause backtracking during search.
  
\item\label{heur:subtyping} Subtyping is only invoked in three places:
  (a) for switching from checking to inference mode (rule
  \textbf{alg-r-}$\uparrow\downarrow$ in
  \Cref{fig:selected-alg-typing-rules}), (b) for the algorithmic
  version of the \textbf{nochange} rule
  (\Cref{fig:synch-rel-typing-rules}), which checks subtyping on all
  variables in the context, and (c) as mentioned in the next point.
  
\item \label{heur:box}
  Relational subtyping rules that mention
  $\tbox{}$ are applied lazily at specific elimination points. For
  instance, in typing a function application, if the applied
  expression's inferred type is $\tbox{(\tau_1 \tarrd{k} \tau_2)}$, we
  try to complete the typing by subtyping to $\tbox{\tau_1} \tarrd{0}
  \tbox{\tau_2}$ and $\tau_1 \tarrd{k} \tau_2$, in that order.
  
\item We switch to the unary reasoning (algorithmic analogues of rule
  \textbf{switch} from \Cref{fig:synch-rel-typing-rules}) only when
  necessary \ie\ when (a) eliminating expressions of the type
  $\tch{\grt_1}{\grt_2}$, (b) checking related expressions at type
  $\tch{\grt_1}{\grt_2}$, and (c) no other relational rules apply
  (this happens when the expressions being related are structurally
  dissimilar at the top-level).
  
\end{enumerate}

These heuristics suffice for all the examples we have seen so far. We
list some of these examples later.

\paragraph{Constraint solving}
In principle, we could pass the constraints \textcolor{red}{$\Phi$}
output by {\bitname}'s checking and inference judgments to an SMT
solver that understands the domain of integers (for sizes) and real
numbers (for costs). However, the constraint typically contains many
existentially quantified variables and current SMT solvers do not
eliminate such variables well. To solve this problem, we wrote a
simple pre-processing pass that finds candidate substitutions for
existentially quantified variables. For any such variable $n$, we look
for constraints of the form $n = I$ and $n \leq I$. In either case, we
consider $I$ a candidate substitution for $n$. In this way, we
generate a set of candidate substitutions for all existentially
quantified variables. For each such substitution, we try to check the
constraint's satisfiability using an SMT solver. (Our implementation
actually does this lazily: It generates a candidate substitution;
calls SMT; if that fails, generates the next substitution, and so on.)


To check the satisfiability of existential-free constraints, we invoke
an SMT solver. Specifically, we use Why3~\cite{Filliatre:2013}, a
common front-end for many SMT solvers. Empirically, we have observed
that only one SMT solver, Alt-Ergo~\cite{bobot2013alt}, can handle our
constraints and, so, our implementation uses this solver behind
Why3. Why3 provides libraries of lemmas for exponentiation, logarithms
and iterated sums, which we use in some of the examples. For typing
programs that use divide-and-conquer over lists (e.g., merge sort), we
have to provide as an axiom one additional lemma that solves a general
recurrence related to costs. This lemma was proved in prior work
(Lemma 2 in the appendix of~\cite{CostIt15}).

\paragraph{Experimental evaluation}
We have used our implementation to typecheck a variety of examples,
including all the examples from the {\tname} paper. Some of the
examples, such as the relational analysis of merge sort
($\kw{msort}$), have rather complex paper proofs. However, in all
cases, the total typechecking time (including existential elimination
and SMT solving) is less than 1s, suggesting that the approach is
practical. Table~\ref{tab:benchmarks} shows the experimental results
over a subset of our example programs (our appendix lists all our
example programs, including their code and experimental results).  A
``-'' indicates a negligible value. Our experiments were performed on
a 3.20GHz 4-core Intel Core i5-6500 processor with 16 GB of RAM.

We briefly describe some of the example programs in
Table~\ref{tab:benchmarks} to highlight their diversity. The program
$\kw{map}$ is the list map function of Section~\ref{sec:relref}. The
program $\kw{comp}$ is a constant-time ($0$ relative cost) comparison
function that checks the equality of two passwords, represented as
lists of bits. The program $\kw{sam}$ (square-and-multiply) computes
the positive powers of a number, represented as a list of bits. The
experiment $\kw{find}$ compares two functions that find a given
element by scanning a list from head to tail and tail to head,
respectively. The program $\kw{2Dcount}$ counts the number of rows of
a matrix, represented as a list of lists in row-major form, that
satisfy a predicate $p$ and contain a key $x$. The program
$\kw{bsplit}$ splits a list into two nearly equal length lists. The
program $\kw{merge}$ merges two sorted lists and the program
$\kw{msort}$ is the standard merge sort function. In all cases except
$\kw{find}$, the goal of the analysis is to find an upper bound on the
relative cost of the function as its input changes. In all cases, the
cost bounds we check are asymptotically tight.

\begin{table}[t]
  \centering
  \begin{tabular} {| p{1.6cm} | p{1.0cm} | p{1.2cm} | p{1.5cm} |
      p{1.5cm} |}
    \hline
    Benchmark        & Total time(s) & Type-checking & Existential elim.  & Constraint solving \\ \hline
 $\kw{filter}$ & 0.13 & - & - & 0.13 \\ \hline 
 $\kw{append}$ & 0.14 & - & - & 0.13 \\ \hline 
 $\kw{rev}$ & 0.15 & - & - & 0.15 \\ \hline 
$\kw{map}$ & 0.11 & - & - & 0.11 \\ \hline 
$\kw{comp}$ & 0.08 & - & - & 0.07 \\ \hline 
$\kw{sam}$ & 0.09 & 0.01 & - & 0.08 \\ \hline 
$\kw{find}$ & 0.05 & - & - & 0.04 \\ \hline 
$\kw{2Dcount}$ & 0.06 & - & - & 0.06 \\ \hline 
 $\kw{ssort}$ & 0.07 & - & - & 0.07 \\ \hline 
$\kw{bsplit}$ & 0.17 & - & - & 0.17 \\ \hline 
 $\kw{flatten}$ & 0.06 & - & - & 0.05 \\ \hline 
 $\kw{appSum}$ & 0.10 & - & - & 0.09 \\ \hline 
$\kw{merge}$ & 0.12 & - & - & 0.12 \\ \hline 
 $\kw{zip}$ & 0.13 & - & - & 0.13 \\ \hline 
$\kw{msort}$ & 0.40 & 0.01 & 0.02 & 0.36 \\ \hline 
 $\kw{bfold}$ & 0.77 & - & 0.01 & 0.77 \\ \hline 

  \end{tabular}
  \caption{{\bitname} runtime on benchmarks from the {\tname} paper
    and the function $\kw{map}$. All times are in seconds.}
  \label{tab:benchmarks}
\end{table}

\paragraph{Annotation effort} 
In a traditional bidirectional type system, the programmer's
annotation effort is limited to providing the eliminated type at every
explicit $\beta$-redex and the type of every top-level function
definition in the program. In our setting, the burden is similar,
except that type annotations on functions also include a cost (on the
arrow). In all but one of the examples we have tried, annotations are
only necessary at each top-level function. One example has an explicit
$\beta$-redex (in the form of a let-binding) and needs one additional
annotation. Overall, this shows that the bidirectional approach
extends to the relational setting (even with refinements and effects),
without adding annotation burden.

\paragraph{Heuristics illustrated with merge sort}
We explain how our implementation types one example---the standard
merge sort function--- \textit{relationally}. The goal of this
exercise is primarily to illustrate some of our heuristics.  The merge
sort function, $\kw{msort}$, splits a list into two nearly equal-sized
sublists using an auxiliary function we call $\kw{bsplit}$,
recursively sorts each sublist and then merges the two sorted sublists
using another function $\kw{merge}$. In their paper on {\tname},
\citet{Cicek:2017} show that the relative cost of two runs of
$\kw{msort}$ with two input lists of length $n$ that differ in at most
$\alpha$ positions is upper-bounded by $Q(n,\alpha) =
\ssum{i}{\szero}{H}{\smult{h(\sceil{\sdiv{2^i}{2}})}{\smin{\alpha}{2^{\sminus{H}{i}}}}}$,
where $H=\sceil{\slog{n}}$ and $h$ is a specific linear function. This
open-form expression lies in $O(n\cdot(1 +
\slog{\alpha}))$.\footnote{This relative cost $O(n\cdot(1 +
  \slog{\alpha}))$ is asymptotically better than the relative cost
  $O(n\cdot\slog{n})$ that can be established non-relationally. Also,
  this relative cost bound is tight.}  Next, we explain at a
high-level how this relative cost is typechecked bidirectionally. We
show below the code of the top-level merge sort function $\kw{msort}$.
\begin{tabbing}
$\efix \kw{msort}(\_). \eLam. \eLam. \elam l. \caseof{l}$
$\ofnil{\enil}$\\
 $\ofcons{$\=$h_1}{tl_1}{} \caseof{tl_1}$\=$\ofnil{\econs(h_1,\enil)}$\\
\>$\ofcons{\_}{\_}{$\=$\elet r = \kw{bsplit} \eapp () \eApp \eApp \eapp l \ein$\\
  \>\> $ \eunpack r \eas r' \ein $$ \clet r' \eas (z_1,z_2) $\\
  \> $\ein \kw{merge} \eapp () \eApp \eApp \eapp (\kw{msort} \eapp () \eApp \eApp \eapp z_1, \kw{msort} \eapp () \eApp \eApp \eapp z_2)}$
\end{tabbing}
We do not show the code of the helper functions $\kw{bsplit}$ and
$\kw{merge}$, but they have the following types (these types are also
checked with {\bitname}; we omit those details here):
\[
  \begin{array}{l@{}l}
    \kw{bsplit}:~&\tbox{(\trunit \to \tforallN{n,\alpha}
    \tlist{n}{\alpha}{\tau} \to \\ 
    &\texistsN{\beta} \tcprod{\beta \leq
    \alpha}{(\tlist{\sceil{\sdiv{n}{2}}}{\beta}{\tau} \ttimes
    \tlist{\sfloor{\sdiv{n}{2}}}{\sminus{\alpha}{\beta}}{\tau})})}
\end{array}
\]
\[
  \begin{array}{l@{}l}
    \kw{merge} : ~&
    \tbox{(\tcho{(\tunit \to \tforallN{n,m} (\ulist{n}{\tint} \ttimes
    \ulist{m}{\tint})  \\
   &\uarr{h(\smin{n}{m})}{h(n+m)}
    \ulist{n+m}{\tint})})}
   \end{array}
\]
Note the $\tbox{}$-es outside the types; their significance will be
clear soon. Our aim is to typecheck $\kw{msort}$ relative to itself at
the following type:
\[\tbox{(\trunit \to \tforallN{n,\alpha} \tlist{n}{\alpha}{(\tcho{\tint})}
    \tarrd{Q(n,\alpha)} \tcho{(\ulist{n}{\tint})})}
\]  
Here, $Q(n,\alpha)$ is the cost function defined above.  We focus on
the most interesting part where we call $\kw{merge}$ on the results of
the two recursive calls to $\kw{msort}$. At this point, we have $z_1:
\tlist{\sceil{\sdiv{n}{2}}}{\beta}{(\tcho{\tint})}$ and $z_2:
\tlist{\sfloor{\sdiv{n}{2}}}{\alpha-\beta}{(\tcho{\tint})}$ in the
context (from the type of $\kw{bsplit}$). Considering that only the
calls to $\kw{merge}$ and $\kw{msort}$ incur additional costs (all the
remaining operations occur synchronously on both sides and the
relative cost of $\kw{bsplit}$ is $0$ from its type), if we were to
naively establish the bound $Q(n,\alpha)$, we would have to show the
following inequality:
\begin{equation}
h(\sceil{\sdiv{n}{2}}) + Q(\sceil{\sdiv{n}{2}}, \beta) + Q(\sfloor{\sdiv{n}{2}}, \alpha
-\beta) \leq Q(n,\alpha)
\label{eq:msort:arith}
\end{equation}
where the cost $h(\sceil{\sdiv{n}{2}}) = h(n) -
h(min(\sceil{\sdiv{n}{2}},\sfloor{\sdiv{n}{2}}))$ comes from the
relative cost of $\kw{merge}$. However, as observed in the {\tname}
paper, this inequality holds only when $\alpha > 0$. When $\alpha=0$,
the right hand side is $0$ whereas the left hand side is
$h(\sceil{\sdiv{n}{2}})$. Nevertheless, when $\alpha=0$, the two input
lists do not differ at all, so the relative cost of $\kw{merge}$ can
be trivially established as $0$ using the \textbf{nochange} rule.

Consequently, the verification of $\kw{merge}$'s body differs based on
whether $\alpha=0$ or not. In our implementation, this case analysis
is provided for by heuristic (\ref{heur:split}). As soon as the list
$l$ is introduced into the context, we apply the algorithmic analogue
of the rule \textbf{rr-split}, introducing the two cases
$\ceq{\alpha}{0}$ and $\alpha >0$. For the case $\ceq{\alpha}{0}$, the
heuristic immediately invokes the rule
\textbf{alg-r-nochange-$\downarrow$}, which requires us to show that
all the free variables in the context have $\square$-ed types. Since
we know that the functions $\kw{merge}$, $\kw{bsplit}$ and
$\kw{msort}$ all have $\square$-ed types, what remains to be shown is
that $\tlist{n}{\alpha}{\tau_1} \sqsubseteq
\tbox{(\tlist{n}{\alpha}{\tau_1})}$. At this point,
heuristic~(\ref{heur:subtyping}) kicks in and uses the subtyping rules
\textbf{l2}, \textbf{l$\tbox{}$} (\Cref{fig:relref-ts}) and the
case-constraint $\ceq{\alpha}{0}$ to establish this property.  For the
other case ($\alpha > 0$), we type the function body following its
syntax. This eventually generates the satisfiable constraint
(\ref{eq:msort:arith}) that is discharged by the constraint
solver. The verification also uses heuristic~(\ref{heur:box}) to
typecheck the applications of $\kw{bsplit}$ and $\kw{merge}$ despite
their $\tbox{}$-ed types.



\section{Related Work}
\label{sec:related}

There is a lot of literature on implementing various combinations of
refinement types, effect systems, modal types, and subtyping. A
distinctive feature of our work is that it combines all these
aspects in a relational setting.

The idea of bidirectional type
systems appeared in literature early on. However, the idea was
popularized only more recently by Pierce and
Turner~\cite{Pierce:2000}. The technique has shown great
applicability---it has been used for dependent
types~\cite{Coquand:96}, indexed refinement
types~\cite{xi99:dependent,XiThesis}, intersection and union
types~\cite{Davies:2000,Dunfield:2003}, higher-rank
polymorphism~\cite{PeytonJones:2007,Dunfield13:bidir,DunfieldK16},
contextual modal types~\cite{Pientka:2008}, algebraic effect
handlers~\cite{Lindley:2017} and gradual typing~\cite{Toro:2015}. Our
approach is inspired by many of these papers, in particular
DML~\cite{xi99:dependent,XiThesis}, but departs in the technical
design of the algorithmic type system due to new challenges offered by
relational and modal types, and costs. In particular, in 
all these works 
the reasoning principle is \textit{unary}, i.e.\ a single program is
checked (inferred) in isolation. Moreover, none of these works
consider effects explicitly, i.e.\ as a type and effect system. One
exception is the bidirectional effect system by ~\cite{Toro:2015},
which uses bidirectional typechecking for gradual unary
effects. However, their end goal is different since they infer minimal
effects at compile time and then check dynamic effects at runtime.


Numerous other systems use lightweight
dependent types for program verification including, for instance,
F$^*$~\cite{SwamyCFSBY11,SwamyHKRDFBFSKZ16} and
LiquidHaskell~\cite{VazouSJ14}. However, these developments also do
not consider comonadic types and costs.
The DML approach has also been used in combination
with linear types for asymptotic complexity analysis
\cite{lago12,Dallago:13}
 and for reasoning about differential
privacy~\cite{gaboardi13:dep, AmorimGAH14}. Besides lightweight dependent types, these
papers also consider the comonadic modality of linear logic. This
modality's structural properties are quite different from those of the
comonadic  
$\tbox{}$ we consider here. 
Another way to extend dependent types with cost information is to index a
monad with the execution cost. Gundry considers this approach in a unary setting for a
subset of Haskell with support for bidirectional
typechecking~\cite{GundryThesis}. 
None of these papers consider
relational typing.



Some other type systems
establish relational properties of programs. For
instance,~\citet{BFGSSZ14} consider a relational variant of a fragment
of F$^*$ for the verification of cryptographic implementations, and
similarly \citet{BartheGAHRS15} consider a relational refinement type
system for differential privacy. However, some of the key technical
challenges of our system, including those that arise from the
interaction between unary and relational typing, as well as costs, do
not show up in these settings. Moreover, these systems use
verification condition generation, not bidirectionality.
In the realm of incremental computing, some
work~\cite{CostIt15,DuCostIt16} has proposed declarative type systems
for reasoning about update costs of incremental programs. These
systems share similarities with the systems we considered here and we
believe that the ideas developed in this paper can be applied to
obtain algorithmic versions of these type systems as well.

Prior work has also studied
methods of eliminating subtyping as a way of simplifying type
checking, e.g.~\cite{Breazu-Tannen:1991,Crary:2000}.  While our
approach is similar in motivation, our technical challenges are quite
different. Main difficulties in simplifying subtyping in our work
arise from the interaction of the modalities $\tbox{}$ and $U$ with
other connectives.


\section{Conclusion}
\label{sec:conclusion}
%
This paper presented a theoretical study and a concrete implementation
of bidirectional type checking in a setting that combines relational
refinements, comonadic types and relational effects. This rich setting
poses unique challenges: The typing rules are not syntax-directed due
to relational refinements; switching from relational to unary
reasoning adds to the ambiguity; subtyping for (relational) comonads
poses additional problems, as do the relational effects. We resolve
these challenges through a process of elaboration and
subtyping-elimination in the theory and using example-guided
heuristics in the implementation. We validate experimentally that this
approach is practical---it works for many different kinds of programs,
has little annotation burden and typechecking is quick. Although we
have focused here on a specific line of type systems with the features
mentioned above, we believe that our work will help future designers
of other relational type (and effect) systems as well.


\bibliography{main}


\begin{thebibliography}{45}


\ifx \showCODEN    \undefined \def \showCODEN     #1{\unskip}     \fi
\ifx \showDOI      \undefined \def \showDOI       #1{#1}\fi
\ifx \showISBNx    \undefined \def \showISBNx     #1{\unskip}     \fi
\ifx \showISBNxiii \undefined \def \showISBNxiii  #1{\unskip}     \fi
\ifx \showISSN     \undefined \def \showISSN      #1{\unskip}     \fi
\ifx \showLCCN     \undefined \def \showLCCN      #1{\unskip}     \fi
\ifx \shownote     \undefined \def \shownote      #1{#1}          \fi
\ifx \showarticletitle \undefined \def \showarticletitle #1{#1}   \fi
\ifx \showURL      \undefined \def \showURL       {\relax}        \fi
\providecommand\bibfield[2]{#2}
\providecommand\bibinfo[2]{#2}
\providecommand\natexlab[1]{#1}
\providecommand\showeprint[2][]{arXiv:#2}

\bibitem[\protect\citeauthoryear{Abadi, Cardelli, and Curien}{Abadi
  et~al\mbox{.}}{1993}]%
        {AbadiCC93}
\bibfield{author}{\bibinfo{person}{Mart{\'{\i}}n Abadi}, \bibinfo{person}{Luca
  Cardelli}, {and} \bibinfo{person}{Pierre{-}Louis Curien}.}
  \bibinfo{year}{1993}\natexlab{}.
\newblock \showarticletitle{Formal Parametric Polymorphism}. In
  \bibinfo{booktitle}{\emph{Conference Record of the Twentieth Annual {ACM}
  {SIGPLAN-SIGACT} Symposium on Principles of Programming Languages,
  Charleston, South Carolina, USA, January 1993}}. \bibinfo{pages}{157--170}.
\newblock
\urldef\tempurl%
\url{https://doi.org/10.1145/158511.158622}
\showDOI{\tempurl}


\bibitem[\protect\citeauthoryear{Abel, Vezzosi, and Winterhalter}{Abel
  et~al\mbox{.}}{2017}]%
        {Abel17icfp}
\bibfield{author}{\bibinfo{person}{Andreas Abel}, \bibinfo{person}{Andrea
  Vezzosi}, {and} \bibinfo{person}{Th{\'{e}}o Winterhalter}.}
  \bibinfo{year}{2017}\natexlab{}.
\newblock \showarticletitle{Normalization by evaluation for sized dependent
  types}.
\newblock \bibinfo{journal}{\emph{{PACMPL}}} \bibinfo{volume}{1},
  \bibinfo{number}{{ICFP}} (\bibinfo{year}{2017}),
  \bibinfo{pages}{33:1--33:30}.
\newblock
\urldef\tempurl%
\url{https://doi.org/10.1145/3110277}
\showDOI{\tempurl}


\bibitem[\protect\citeauthoryear{Aguirre, Barthe, Birkedal, Bizjak, Gaboardi,
  and Garg}{Aguirre et~al\mbox{.}}{2018}]%
        {Aguirre18esop}
\bibfield{author}{\bibinfo{person}{Alejandro Aguirre}, \bibinfo{person}{Gilles
  Barthe}, \bibinfo{person}{Lars Birkedal}, \bibinfo{person}{Ales Bizjak},
  \bibinfo{person}{Marco Gaboardi}, {and} \bibinfo{person}{Deepak Garg}.}
  \bibinfo{year}{2018}\natexlab{}.
\newblock \showarticletitle{Relational Reasoning for Markov Chains in a
  Probabilistic Guarded Lambda Calculus}. In
  \bibinfo{booktitle}{\emph{Programming Languages and Systems - 27th European
  Symposium on Programming, {ESOP} 2018, Held as Part of the European Joint
  Conferences on Theory and Practice of Software, {ETAPS} 2018, Thessaloniki,
  Greece, April 14-20, 2018, Proceedings}}. \bibinfo{pages}{214--241}.
\newblock
\urldef\tempurl%
\url{https://doi.org/10.1007/978-3-319-89884-1\_8}
\showDOI{\tempurl}


\bibitem[\protect\citeauthoryear{Aguirre, Barthe, Gaboardi, Garg, and
  Strub}{Aguirre et~al\mbox{.}}{2017}]%
        {AguirreBG0S17}
\bibfield{author}{\bibinfo{person}{Alejandro Aguirre}, \bibinfo{person}{Gilles
  Barthe}, \bibinfo{person}{Marco Gaboardi}, \bibinfo{person}{Deepak Garg},
  {and} \bibinfo{person}{Pierre{-}Yves Strub}.}
  \bibinfo{year}{2017}\natexlab{}.
\newblock \showarticletitle{A relational logic for higher-order programs}.
\newblock \bibinfo{journal}{\emph{{PACMPL}}} \bibinfo{volume}{1},
  \bibinfo{number}{{ICFP}} (\bibinfo{year}{2017}),
  \bibinfo{pages}{21:1--21:29}.
\newblock
\urldef\tempurl%
\url{https://doi.org/10.1145/3110265}
\showDOI{\tempurl}


\bibitem[\protect\citeauthoryear{Barthe, Fournet, Gr{\'{e}}goire, Strub, Swamy,
  and B{\'{e}}guelin}{Barthe et~al\mbox{.}}{2014a}]%
        {BartheFGSSB14}
\bibfield{author}{\bibinfo{person}{Gilles Barthe},
  \bibinfo{person}{C{\'{e}}dric Fournet}, \bibinfo{person}{Benjamin
  Gr{\'{e}}goire}, \bibinfo{person}{Pierre{-}Yves Strub},
  \bibinfo{person}{Nikhil Swamy}, {and} \bibinfo{person}{Santiago~Zanella
  B{\'{e}}guelin}.} \bibinfo{year}{2014}\natexlab{a}.
\newblock \showarticletitle{Probabilistic relational verification for
  cryptographic implementations}. In \bibinfo{booktitle}{\emph{The 41st Annual
  {ACM} {SIGPLAN-SIGACT} Symposium on Principles of Programming Languages,
  {POPL} '14, San Diego, CA, USA, January 20-21, 2014}}.
  \bibinfo{pages}{193--206}.
\newblock
\urldef\tempurl%
\url{https://doi.org/10.1145/2535838.2535847}
\showDOI{\tempurl}


\bibitem[\protect\citeauthoryear{Barthe, Fournet, Gr{\'{e}}goire, Strub, Swamy,
  and B{\'{e}}guelin}{Barthe et~al\mbox{.}}{2014b}]%
        {BFGSSZ14}
\bibfield{author}{\bibinfo{person}{Gilles Barthe},
  \bibinfo{person}{C{\'{e}}dric Fournet}, \bibinfo{person}{Benjamin
  Gr{\'{e}}goire}, \bibinfo{person}{Pierre{-}Yves Strub},
  \bibinfo{person}{Nikhil Swamy}, {and} \bibinfo{person}{Santiago~Zanella
  B{\'{e}}guelin}.} \bibinfo{year}{2014}\natexlab{b}.
\newblock \showarticletitle{Probabilistic relational verification for
  cryptographic implementations}. In \bibinfo{booktitle}{\emph{Proceedings of
  the 41st Annual {ACM} {SIGPLAN-SIGACT} Symposium on Principles of Programming
  Languages, {POPL}'14}}, \bibfield{editor}{\bibinfo{person}{Suresh
  Jagannathan} {and} \bibinfo{person}{Peter Sewell}} (Eds.).
  \bibinfo{pages}{193--206}.
\newblock


\bibitem[\protect\citeauthoryear{Barthe, Gaboardi, Arias, Hsu, Roth, and
  Strub}{Barthe et~al\mbox{.}}{2015a}]%
        {BartheGAHRS15}
\bibfield{author}{\bibinfo{person}{Gilles Barthe}, \bibinfo{person}{Marco
  Gaboardi}, \bibinfo{person}{Emilio Jes{\'{u}}s~Gallego Arias},
  \bibinfo{person}{Justin Hsu}, \bibinfo{person}{Aaron Roth}, {and}
  \bibinfo{person}{Pierre{-}Yves Strub}.} \bibinfo{year}{2015}\natexlab{a}.
\newblock \showarticletitle{Higher-Order Approximate Relational Refinement
  Types for Mechanism Design and Differential Privacy}. In
  \bibinfo{booktitle}{\emph{Proceedings of the 42nd Annual {ACM}
  {SIGPLAN-SIGACT} Symposium on Principles of Programming Languages, {POPL}
  2015, Mumbai, India, January 15-17, 2015}}. \bibinfo{pages}{55--68}.
\newblock
\urldef\tempurl%
\url{https://doi.org/10.1145/2676726.2677000}
\showDOI{\tempurl}


\bibitem[\protect\citeauthoryear{Barthe, Gaboardi, Gallego~Arias, Hsu, Roth,
  and Strub}{Barthe et~al\mbox{.}}{2015b}]%
        {BGGHRS15}
\bibfield{author}{\bibinfo{person}{Gilles Barthe}, \bibinfo{person}{Marco
  Gaboardi}, \bibinfo{person}{Emilio~Jes{\'u}s Gallego~Arias},
  \bibinfo{person}{Justin Hsu}, \bibinfo{person}{Aaron Roth}, {and}
  \bibinfo{person}{Pierre-Yves Strub}.} \bibinfo{year}{2015}\natexlab{b}.
\newblock \showarticletitle{Higher-order approximate relational refinement
  types for mechanism design and differential privacy}. In
  \bibinfo{booktitle}{\emph{Proceedings of the 42nd Annual {ACM}
  {SIGPLAN-SIGACT} Symposium on Principles of Programming Languages, {POPL}
  2015, Mumbai, India, January 15-17, 2015}},
  \bibfield{editor}{\bibinfo{person}{Sriram~K. Rajamani} {and}
  \bibinfo{person}{David Walker}} (Eds.). \bibinfo{pages}{55--68}.
\newblock


\bibitem[\protect\citeauthoryear{Bierman, Meijer, and Torgersen}{Bierman
  et~al\mbox{.}}{2007}]%
        {BiermanMT07}
\bibfield{author}{\bibinfo{person}{Gavin~M. Bierman}, \bibinfo{person}{Erik
  Meijer}, {and} \bibinfo{person}{Mads Torgersen}.}
  \bibinfo{year}{2007}\natexlab{}.
\newblock \showarticletitle{Lost in translation: formalizing proposed
  extensions to c{\#}}. In \bibinfo{booktitle}{\emph{Proceedings of the 22nd
  Annual {ACM} {SIGPLAN} Conference on Object-Oriented Programming, Systems,
  Languages, and Applications, {OOPSLA} 2007, October 21-25, 2007, Montreal,
  Quebec, Canada}}. \bibinfo{pages}{479--498}.
\newblock
\urldef\tempurl%
\url{https://doi.org/10.1145/1297027.1297063}
\showDOI{\tempurl}


\bibitem[\protect\citeauthoryear{Bobot, Conchon, Contejean, Iguernelala,
  Lescuyer, and Mebsout}{Bobot et~al\mbox{.}}{2013}]%
        {bobot2013alt}
\bibfield{author}{\bibinfo{person}{Fran{\c{c}}ois Bobot},
  \bibinfo{person}{Sylvain Conchon}, \bibinfo{person}{E Contejean},
  \bibinfo{person}{Mohamed Iguernelala}, \bibinfo{person}{St{\'e}phane
  Lescuyer}, {and} \bibinfo{person}{Alain Mebsout}.}
  \bibinfo{year}{2013}\natexlab{}.
\newblock \bibinfo{title}{The Alt-Ergo automated theorem prover, 2008}.
\newblock
\newblock


\bibitem[\protect\citeauthoryear{Bracha, Odersky, Stoutamire, and
  Wadler}{Bracha et~al\mbox{.}}{1998}]%
        {BrachaOSW98}
\bibfield{author}{\bibinfo{person}{Gilad Bracha}, \bibinfo{person}{Martin
  Odersky}, \bibinfo{person}{David Stoutamire}, {and} \bibinfo{person}{Philip
  Wadler}.} \bibinfo{year}{1998}\natexlab{}.
\newblock \showarticletitle{Making the Future Safe for the Past: Adding
  Genericity to the Java Programming Language}. In
  \bibinfo{booktitle}{\emph{Proceedings of the 1998 {ACM} {SIGPLAN} Conference
  on Object-Oriented Programming Systems, Languages {\&} Applications {(OOPSLA}
  '98), Vancouver, British Columbia, Canada, October 18-22, 1998.}}
  \bibinfo{pages}{183--200}.
\newblock
\urldef\tempurl%
\url{https://doi.org/10.1145/286936.286957}
\showDOI{\tempurl}


\bibitem[\protect\citeauthoryear{Breazu-Tannen, Coquand, Gunter, and
  Scedrov}{Breazu-Tannen et~al\mbox{.}}{1991}]%
        {Breazu-Tannen:1991}
\bibfield{author}{\bibinfo{person}{Val Breazu-Tannen}, \bibinfo{person}{Thierry
  Coquand}, \bibinfo{person}{Carl~A. Gunter}, {and} \bibinfo{person}{Andre
  Scedrov}.} \bibinfo{year}{1991}\natexlab{}.
\newblock \showarticletitle{Inheritance As Implicit Coercion}.
\newblock \bibinfo{journal}{\emph{Inf. Comput.}} \bibinfo{volume}{93},
  \bibinfo{number}{1} (\bibinfo{date}{July} \bibinfo{year}{1991}),
  \bibinfo{pages}{172--221}.
\newblock


\bibitem[\protect\citeauthoryear{Brunel, Gaboardi, Mazza, and Zdancewic}{Brunel
  et~al\mbox{.}}{2014}]%
        {BrunelGMZ14}
\bibfield{author}{\bibinfo{person}{Alo{\"{\i}}s Brunel}, \bibinfo{person}{Marco
  Gaboardi}, \bibinfo{person}{Damiano Mazza}, {and} \bibinfo{person}{Steve
  Zdancewic}.} \bibinfo{year}{2014}\natexlab{}.
\newblock \showarticletitle{A Core Quantitative Coeffect Calculus}. In
  \bibinfo{booktitle}{\emph{Programming Languages and Systems - 23rd European
  Symposium on Programming, {ESOP} 2014, Held as Part of the European Joint
  Conferences on Theory and Practice of Software, {ETAPS} 2014, Grenoble,
  France, April 5-13, 2014, Proceedings}}. \bibinfo{pages}{351--370}.
\newblock
\urldef\tempurl%
\url{https://doi.org/10.1007/978-3-642-54833-8\_19}
\showDOI{\tempurl}


\bibitem[\protect\citeauthoryear{\c{C}i\c{c}ek, Barthe, Gaboardi, Garg, and
  Hoffmann}{\c{C}i\c{c}ek et~al\mbox{.}}{2017}]%
        {Cicek:2017}
\bibfield{author}{\bibinfo{person}{Ezgi \c{C}i\c{c}ek}, \bibinfo{person}{Gilles
  Barthe}, \bibinfo{person}{Marco Gaboardi}, \bibinfo{person}{Deepak Garg},
  {and} \bibinfo{person}{Jan Hoffmann}.} \bibinfo{year}{2017}\natexlab{}.
\newblock \showarticletitle{Relational Cost Analysis}. In
  \bibinfo{booktitle}{\emph{Proceedings of the 44th ACM SIGPLAN Symposium on
  Principles of Programming Languages}} \emph{(\bibinfo{series}{POPL 2017})}.
  \bibinfo{publisher}{ACM}, \bibinfo{pages}{316--329}.
\newblock


\bibitem[\protect\citeauthoryear{{\c{C}}i{\c{c}}ek, Garg, and
  Acar}{{\c{C}}i{\c{c}}ek et~al\mbox{.}}{2015}]%
        {CostIt15}
\bibfield{author}{\bibinfo{person}{Ezgi {\c{C}}i{\c{c}}ek},
  \bibinfo{person}{Deepak Garg}, {and} \bibinfo{person}{Umut~A. Acar}.}
  \bibinfo{year}{2015}\natexlab{}.
\newblock \showarticletitle{Refinement Types for Incremental Computational
  Complexity}. In \bibinfo{booktitle}{\emph{Programming Languages and Systems -
  24th European Symposium on Programming, {ESOP} 2015, London, UK, April 11-18,
  2015. Proceedings}}. \bibinfo{pages}{406--431}.
\newblock


\bibitem[\protect\citeauthoryear{{\c{C}}i{\c{c}}ek, Paraskevopoulou, and
  Garg}{{\c{C}}i{\c{c}}ek et~al\mbox{.}}{2016}]%
        {DuCostIt16}
\bibfield{author}{\bibinfo{person}{Ezgi {\c{C}}i{\c{c}}ek},
  \bibinfo{person}{Zoe Paraskevopoulou}, {and} \bibinfo{person}{Deepak Garg}.}
  \bibinfo{year}{2016}\natexlab{}.
\newblock \showarticletitle{A Type Theory for Incremental Computational
  Complexity With Control Flow Changes}. In
  \bibinfo{booktitle}{\emph{Proceedings of the 21st International Conference on
  Functional Programming}} \emph{(\bibinfo{series}{ICFP '16})}.
\newblock


\bibitem[\protect\citeauthoryear{Coquand}{Coquand}{1996}]%
        {Coquand:96}
\bibfield{author}{\bibinfo{person}{Thierry Coquand}.}
  \bibinfo{year}{1996}\natexlab{}.
\newblock \showarticletitle{An algorithm for type-checking dependent types}.
\newblock \bibinfo{journal}{\emph{Science of Computer Programming}}
  \bibinfo{volume}{26}, \bibinfo{number}{1} (\bibinfo{year}{1996}),
  \bibinfo{pages}{167 -- 177}.
\newblock


\bibitem[\protect\citeauthoryear{Crary}{Crary}{2000}]%
        {Crary:2000}
\bibfield{author}{\bibinfo{person}{Karl Crary}.}
  \bibinfo{year}{2000}\natexlab{}.
\newblock \showarticletitle{Typed Compilation of Inclusive Subtyping}. In
  \bibinfo{booktitle}{\emph{Proceedings of the Fifth ACM SIGPLAN International
  Conference on Functional Programming}} \emph{(\bibinfo{series}{ICFP '00})}.
  \bibinfo{publisher}{ACM}, \bibinfo{pages}{68--81}.
\newblock
\showISBNx{1-58113-202-6}


\bibitem[\protect\citeauthoryear{Dal~Lago and Gaboardi}{Dal~Lago and
  Gaboardi}{2011}]%
        {lago12}
\bibfield{author}{\bibinfo{person}{Ugo Dal~Lago} {and} \bibinfo{person}{Marco
  Gaboardi}.} \bibinfo{year}{2011}\natexlab{}.
\newblock \showarticletitle{Linear Dependent Types and Relative Completeness}.
  In \bibinfo{booktitle}{\emph{Proceedings of the 2011 IEEE 26th Annual
  Symposium on Logic in Computer Science}} \emph{(\bibinfo{series}{LICS '11})}.
  \bibinfo{pages}{133--142}.
\newblock


\bibitem[\protect\citeauthoryear{Dal~Lago and Petit}{Dal~Lago and
  Petit}{2013}]%
        {Dallago:13}
\bibfield{author}{\bibinfo{person}{Ugo Dal~Lago} {and} \bibinfo{person}{Barbara
  Petit}.} \bibinfo{year}{2013}\natexlab{}.
\newblock \showarticletitle{The Geometry of Types}. In
  \bibinfo{booktitle}{\emph{Proceedings of the 40th Annual Symposium on
  Principles of Programming Languages}} \emph{(\bibinfo{series}{POPL '13})}.
  \bibinfo{pages}{167--178}.
\newblock


\bibitem[\protect\citeauthoryear{Davies and Pfenning}{Davies and
  Pfenning}{2000}]%
        {Davies:2000}
\bibfield{author}{\bibinfo{person}{Rowan Davies} {and} \bibinfo{person}{Frank
  Pfenning}.} \bibinfo{year}{2000}\natexlab{}.
\newblock \showarticletitle{Intersection Types and Computational Effects}. In
  \bibinfo{booktitle}{\emph{Proceedings of the Fifth ACM SIGPLAN International
  Conference on Functional Programming}} \emph{(\bibinfo{series}{ICFP '00})}.
  \bibinfo{pages}{198--208}.
\newblock


\bibitem[\protect\citeauthoryear{de~Amorim, Gaboardi, Arias, and Hsu}{de~Amorim
  et~al\mbox{.}}{2014}]%
        {AmorimGAH14}
\bibfield{author}{\bibinfo{person}{Arthur~Azevedo de Amorim},
  \bibinfo{person}{Marco Gaboardi}, \bibinfo{person}{Emilio Jes{\'{u}}s~Gallego
  Arias}, {and} \bibinfo{person}{Justin Hsu}.} \bibinfo{year}{2014}\natexlab{}.
\newblock \showarticletitle{Really Natural Linear Indexed Type Checking}. In
  \bibinfo{booktitle}{\emph{Proceedings of the 26th 2014 International
  Symposium on Implementation and Application of Functional Languages, {IFL}
  '14, Boston, MA, USA, October 1-3, 2014}}. \bibinfo{pages}{5:1--5:12}.
\newblock


\bibitem[\protect\citeauthoryear{Dunfield and Krishnaswami}{Dunfield and
  Krishnaswami}{2013}]%
        {Dunfield13:bidir}
\bibfield{author}{\bibinfo{person}{Joshua Dunfield} {and}
  \bibinfo{person}{Neelakantan~R. Krishnaswami}.}
  \bibinfo{year}{2013}\natexlab{}.
\newblock \showarticletitle{Complete and Easy Bidirectional Typechecking for
  Higher-Rank Polymorphism}. In \bibinfo{booktitle}{\emph{International
  Conference on Functional Programming}}.
\newblock
\newblock
\shownote{\url{arXiv:1306.6032 [cs.PL]}.}


\bibitem[\protect\citeauthoryear{Dunfield and Krishnaswami}{Dunfield and
  Krishnaswami}{2016}]%
        {DunfieldK16}
\bibfield{author}{\bibinfo{person}{Joshua Dunfield} {and}
  \bibinfo{person}{Neelakantan~R. Krishnaswami}.}
  \bibinfo{year}{2016}\natexlab{}.
\newblock \showarticletitle{Sound and Complete Bidirectional Typechecking for
  Higher-Rank Polymorphism with Existentials and Indexed Types}.
\newblock \bibinfo{journal}{\emph{CoRR}}  \bibinfo{volume}{abs/1601.05106}
  (\bibinfo{year}{2016}).
\newblock
\urldef\tempurl%
\url{http://arxiv.org/abs/1601.05106}
\showURL{%
\tempurl}


\bibitem[\protect\citeauthoryear{Dunfield and Pfenning}{Dunfield and
  Pfenning}{2003}]%
        {Dunfield:2003}
\bibfield{author}{\bibinfo{person}{Joshua Dunfield} {and}
  \bibinfo{person}{Frank Pfenning}.} \bibinfo{year}{2003}\natexlab{}.
\newblock \showarticletitle{Type Assignment for Intersections and Unions in
  Call-by-value Languages}. In \bibinfo{booktitle}{\emph{Proceedings of the 6th
  International Conference on Foundations of Software Science and Computation
  Structures and Joint European Conference on Theory and Practice of Software}}
  \emph{(\bibinfo{series}{FOSSACS'03/ETAPS'03})}.
  \bibinfo{publisher}{Springer-Verlag}, \bibinfo{pages}{250--266}.
\newblock


\bibitem[\protect\citeauthoryear{Felleisen}{Felleisen}{1991}]%
        {DBLP:journals/scp/Felleisen91}
\bibfield{author}{\bibinfo{person}{Matthias Felleisen}.}
  \bibinfo{year}{1991}\natexlab{}.
\newblock \showarticletitle{On the Expressive Power of Programming Languages}.
\newblock \bibinfo{journal}{\emph{Science of Computer Programming}}
  \bibinfo{volume}{17}, \bibinfo{number}{1-3} (\bibinfo{year}{1991}),
  \bibinfo{pages}{35--75}.
\newblock


\bibitem[\protect\citeauthoryear{Filli\^{a}tre and Paskevich}{Filli\^{a}tre and
  Paskevich}{2013}]%
        {Filliatre:2013}
\bibfield{author}{\bibinfo{person}{Jean-Christophe Filli\^{a}tre} {and}
  \bibinfo{person}{Andrei Paskevich}.} \bibinfo{year}{2013}\natexlab{}.
\newblock \showarticletitle{Why3: Where Programs Meet Provers}. In
  \bibinfo{booktitle}{\emph{Proceedings of the 22Nd European Conference on
  Programming Languages and Systems}} \emph{(\bibinfo{series}{ESOP'13})}.
  \bibinfo{publisher}{Springer-Verlag}, \bibinfo{pages}{125--128}.
\newblock


\bibitem[\protect\citeauthoryear{Gaboardi, Haeberlen, Hsu, Narayan, and
  Pierce}{Gaboardi et~al\mbox{.}}{2013}]%
        {gaboardi13:dep}
\bibfield{author}{\bibinfo{person}{Marco Gaboardi}, \bibinfo{person}{Andreas
  Haeberlen}, \bibinfo{person}{Justin Hsu}, \bibinfo{person}{Arjun Narayan},
  {and} \bibinfo{person}{Benjamin~C. Pierce}.} \bibinfo{year}{2013}\natexlab{}.
\newblock \showarticletitle{Linear Dependent Types for Differential Privacy}.
  In \bibinfo{booktitle}{\emph{Proceedings of the 40th Annual Symposium on
  Principles of Programming Languages}} \emph{(\bibinfo{series}{POPL '13})}.
  \bibinfo{pages}{357--370}.
\newblock


\bibitem[\protect\citeauthoryear{Gundry}{Gundry}{2013}]%
        {GundryThesis}
\bibfield{author}{\bibinfo{person}{Adam Gundry}.}
  \bibinfo{year}{2013}\natexlab{}.
\newblock \emph{\bibinfo{title}{Type Inference, Haskell and Dependent Types}}.
\newblock \bibinfo{thesistype}{Ph.D. Dissertation}. \bibinfo{school}{University
  of Strathclyde}.
\newblock
\newblock
\shownote{available as
  {\url{http://adam.gundry.co.uk/pub/thesis/thesis-2013-12-03.pdf}}.}


\bibitem[\protect\citeauthoryear{Lindley, McBride, and McLaughlin}{Lindley
  et~al\mbox{.}}{2017}]%
        {Lindley:2017}
\bibfield{author}{\bibinfo{person}{Sam Lindley}, \bibinfo{person}{Conor
  McBride}, {and} \bibinfo{person}{Craig McLaughlin}.}
  \bibinfo{year}{2017}\natexlab{}.
\newblock \showarticletitle{Do Be Do Be Do}. In
  \bibinfo{booktitle}{\emph{Proceedings of the 44th ACM SIGPLAN Symposium on
  Principles of Programming Languages}} \emph{(\bibinfo{series}{POPL 2017})}.
  \bibinfo{publisher}{ACM}, \bibinfo{address}{New York, NY, USA},
  \bibinfo{pages}{500--514}.
\newblock


\bibitem[\protect\citeauthoryear{Lucassen and Gifford}{Lucassen and
  Gifford}{1988}]%
        {Lucassen:1988}
\bibfield{author}{\bibinfo{person}{J.~M. Lucassen} {and} \bibinfo{person}{D.~K.
  Gifford}.} \bibinfo{year}{1988}\natexlab{}.
\newblock \showarticletitle{Polymorphic Effect Systems}. In
  \bibinfo{booktitle}{\emph{Proceedings of the 15th ACM SIGPLAN-SIGACT
  Symposium on Principles of Programming Languages}}
  \emph{(\bibinfo{series}{POPL '88})}. \bibinfo{publisher}{ACM},
  \bibinfo{pages}{47--57}.
\newblock


\bibitem[\protect\citeauthoryear{Nielson and Nielson}{Nielson and
  Nielson}{1999}]%
        {nielsen99}
\bibfield{author}{\bibinfo{person}{Flemming Nielson} {and}
  \bibinfo{person}{HanneRiis Nielson}.} \bibinfo{year}{1999}\natexlab{}.
\newblock \showarticletitle{Type and Effect Systems}.
\newblock In \bibinfo{booktitle}{\emph{Correct System Design}}.
  \bibinfo{series}{Lecture Notes in Computer Science},
  Vol.~\bibinfo{volume}{1710}. \bibinfo{publisher}{Springer-Verlag},
  \bibinfo{pages}{114--136}.
\newblock


\bibitem[\protect\citeauthoryear{Odersky, Zenger, and Zenger}{Odersky
  et~al\mbox{.}}{2001}]%
        {ozz:colored-local}
\bibfield{author}{\bibinfo{person}{Martin Odersky}, \bibinfo{person}{Matthias
  Zenger}, {and} \bibinfo{person}{Christoph Zenger}.}
  \bibinfo{year}{2001}\natexlab{}.
\newblock \showarticletitle{Colored Local Type Inference}. In
  \bibinfo{booktitle}{\emph{Proc. ACM Symposium on Principles of Programming
  Languages}}. \bibinfo{pages}{41--53}.
\newblock


\bibitem[\protect\citeauthoryear{Petricek, Orchard, and Mycroft}{Petricek
  et~al\mbox{.}}{2013}]%
        {PetricekOM13}
\bibfield{author}{\bibinfo{person}{Tomas Petricek}, \bibinfo{person}{Dominic~A.
  Orchard}, {and} \bibinfo{person}{Alan Mycroft}.}
  \bibinfo{year}{2013}\natexlab{}.
\newblock \showarticletitle{Coeffects: Unified Static Analysis of
  Context-Dependence}. In \bibinfo{booktitle}{\emph{Automata, Languages, and
  Programming - 40th International Colloquium, {ICALP} 2013, Riga, Latvia, July
  8-12, 2013, Proceedings, Part {II}}}. \bibinfo{pages}{385--397}.
\newblock
\urldef\tempurl%
\url{https://doi.org/10.1007/978-3-642-39212-2\_35}
\showDOI{\tempurl}


\bibitem[\protect\citeauthoryear{Peyton~Jones, Vytiniotis, Weirich, and
  Shields}{Peyton~Jones et~al\mbox{.}}{2007}]%
        {PeytonJones:2007}
\bibfield{author}{\bibinfo{person}{Simon Peyton~Jones},
  \bibinfo{person}{Dimitrios Vytiniotis}, \bibinfo{person}{Stephanie Weirich},
  {and} \bibinfo{person}{Mark Shields}.} \bibinfo{year}{2007}\natexlab{}.
\newblock \showarticletitle{Practical Type Inference for Arbitrary-rank Types}.
\newblock \bibinfo{journal}{\emph{J. Funct. Program.}} \bibinfo{volume}{17},
  \bibinfo{number}{1} (\bibinfo{date}{Jan.} \bibinfo{year}{2007}),
  \bibinfo{pages}{1--82}.
\newblock


\bibitem[\protect\citeauthoryear{{Peyton Jones}, Vytiniotis, Weirich, and
  Shields}{{Peyton Jones} et~al\mbox{.}}{2007}]%
        {JonesVWS07}
\bibfield{author}{\bibinfo{person}{Simon~L. {Peyton Jones}},
  \bibinfo{person}{Dimitrios Vytiniotis}, \bibinfo{person}{Stephanie Weirich},
  {and} \bibinfo{person}{Mark Shields}.} \bibinfo{year}{2007}\natexlab{}.
\newblock \showarticletitle{Practical type inference for arbitrary-rank types}.
\newblock \bibinfo{journal}{\emph{J. Funct. Program.}} \bibinfo{volume}{17},
  \bibinfo{number}{1} (\bibinfo{year}{2007}), \bibinfo{pages}{1--82}.
\newblock
\urldef\tempurl%
\url{https://doi.org/10.1017/S0956796806006034}
\showDOI{\tempurl}


\bibitem[\protect\citeauthoryear{Pientka}{Pientka}{2008}]%
        {Pientka:2008}
\bibfield{author}{\bibinfo{person}{Brigitte Pientka}.}
  \bibinfo{year}{2008}\natexlab{}.
\newblock \showarticletitle{A Type-theoretic Foundation for Programming with
  Higher-order Abstract Syntax and First-class Substitutions}. In
  \bibinfo{booktitle}{\emph{Proceedings of the 35th Annual ACM SIGPLAN-SIGACT
  Symposium on Principles of Programming Languages}}
  \emph{(\bibinfo{series}{POPL '08})}. \bibinfo{pages}{371--382}.
\newblock


\bibitem[\protect\citeauthoryear{Pierce and Turner}{Pierce and Turner}{2000}]%
        {Pierce:2000}
\bibfield{author}{\bibinfo{person}{Benjamin~C. Pierce} {and}
  \bibinfo{person}{David~N. Turner}.} \bibinfo{year}{2000}\natexlab{}.
\newblock \showarticletitle{Local Type Inference}.
\newblock \bibinfo{journal}{\emph{ACM Trans. Program. Lang. Syst.}}
  \bibinfo{volume}{22}, \bibinfo{number}{1} (\bibinfo{date}{Jan.}
  \bibinfo{year}{2000}), \bibinfo{pages}{1--44}.
\newblock
\showISSN{0164-0925}


\bibitem[\protect\citeauthoryear{Pottier and Simonet}{Pottier and
  Simonet}{2003}]%
        {Pottier03:IF}
\bibfield{author}{\bibinfo{person}{Fran\c{c}ois Pottier} {and}
  \bibinfo{person}{Vincent Simonet}.} \bibinfo{year}{2003}\natexlab{}.
\newblock \showarticletitle{Information Flow Inference for {ML}}.
\newblock \bibinfo{journal}{\emph{ACM Trans. Prog. Lang. Sys.}}
  \bibinfo{volume}{25}, \bibinfo{number}{1} (\bibinfo{date}{Jan.}
  \bibinfo{year}{2003}), \bibinfo{pages}{117--158}.
\newblock


\bibitem[\protect\citeauthoryear{Swamy, Chen, Fournet, Strub, Bhargavan, and
  Yang}{Swamy et~al\mbox{.}}{2011}]%
        {SwamyCFSBY11}
\bibfield{author}{\bibinfo{person}{Nikhil Swamy}, \bibinfo{person}{Juan Chen},
  \bibinfo{person}{C{\'{e}}dric Fournet}, \bibinfo{person}{Pierre{-}Yves
  Strub}, \bibinfo{person}{Karthikeyan Bhargavan}, {and} \bibinfo{person}{Jean
  Yang}.} \bibinfo{year}{2011}\natexlab{}.
\newblock \showarticletitle{Secure distributed programming with value-dependent
  types}. In \bibinfo{booktitle}{\emph{Proceeding of the 16th {ACM} {SIGPLAN}
  international conference on Functional Programming, {ICFP} 2011, Tokyo,
  Japan, September 19-21, 2011}}. \bibinfo{pages}{266--278}.
\newblock
\urldef\tempurl%
\url{https://doi.org/10.1145/2034773.2034811}
\showDOI{\tempurl}


\bibitem[\protect\citeauthoryear{Swamy, Hritcu, Keller, Rastogi,
  Delignat{-}Lavaud, Forest, Bhargavan, Fournet, Strub, Kohlweiss,
  Zinzindohoue, and B{\'{e}}guelin}{Swamy et~al\mbox{.}}{2016}]%
        {SwamyHKRDFBFSKZ16}
\bibfield{author}{\bibinfo{person}{Nikhil Swamy}, \bibinfo{person}{Catalin
  Hritcu}, \bibinfo{person}{Chantal Keller}, \bibinfo{person}{Aseem Rastogi},
  \bibinfo{person}{Antoine Delignat{-}Lavaud}, \bibinfo{person}{Simon Forest},
  \bibinfo{person}{Karthikeyan Bhargavan}, \bibinfo{person}{C{\'{e}}dric
  Fournet}, \bibinfo{person}{Pierre{-}Yves Strub}, \bibinfo{person}{Markulf
  Kohlweiss}, \bibinfo{person}{Jean~Karim Zinzindohoue}, {and}
  \bibinfo{person}{Santiago~Zanella B{\'{e}}guelin}.}
  \bibinfo{year}{2016}\natexlab{}.
\newblock \showarticletitle{Dependent types and multi-monadic effects in
  {F}$^*$}. In \bibinfo{booktitle}{\emph{Proceedings of the 43rd Annual {ACM}
  {SIGPLAN-SIGACT} Symposium on Principles of Programming Languages, {POPL}
  2016, St. Petersburg, FL, USA, January 20 - 22, 2016}}.
  \bibinfo{pages}{256--270}.
\newblock
\urldef\tempurl%
\url{https://doi.org/10.1145/2837614.2837655}
\showDOI{\tempurl}


\bibitem[\protect\citeauthoryear{Toro and Tanter}{Toro and Tanter}{2015}]%
        {Toro:2015}
\bibfield{author}{\bibinfo{person}{Mat\'{\i}as Toro} {and}
  \bibinfo{person}{\'{E}ric Tanter}.} \bibinfo{year}{2015}\natexlab{}.
\newblock \showarticletitle{Customizable Gradual Polymorphic Effects for
  Scala}. In \bibinfo{booktitle}{\emph{Proceedings of the 2015 ACM SIGPLAN
  International Conference on Object-Oriented Programming, Systems, Languages,
  and Applications}} \emph{(\bibinfo{series}{OOPSLA 2015})}.
  \bibinfo{publisher}{ACM}, \bibinfo{address}{New York, NY, USA},
  \bibinfo{pages}{935--953}.
\newblock


\bibitem[\protect\citeauthoryear{Vazou, Seidel, and Jhala}{Vazou
  et~al\mbox{.}}{2014}]%
        {VazouSJ14}
\bibfield{author}{\bibinfo{person}{Niki Vazou}, \bibinfo{person}{Eric~L.
  Seidel}, {and} \bibinfo{person}{Ranjit Jhala}.}
  \bibinfo{year}{2014}\natexlab{}.
\newblock \showarticletitle{LiquidHaskell: experience with refinement types in
  the real world}. In \bibinfo{booktitle}{\emph{Proceedings of the 2014 {ACM}
  {SIGPLAN} symposium on Haskell, Gothenburg, Sweden, September 4-5, 2014}}.
  \bibinfo{pages}{39--51}.
\newblock
\urldef\tempurl%
\url{https://doi.org/10.1145/2633357.2633366}
\showDOI{\tempurl}


\bibitem[\protect\citeauthoryear{Xi}{Xi}{1998}]%
        {XiThesis}
\bibfield{author}{\bibinfo{person}{Hongwei Xi}.}
  \bibinfo{year}{1998}\natexlab{}.
\newblock \bibinfo{thesistype}{Ph.D. Dissertation}. \bibinfo{school}{Carnegie
  Mellon University}.
\newblock
\newblock
\shownote{available as {\url{https://www.cs.cmu.edu/~rwh/theses/xi.pdf}}.}


\bibitem[\protect\citeauthoryear{Xi and Pfenning}{Xi and Pfenning}{1999}]%
        {xi99:dependent}
\bibfield{author}{\bibinfo{person}{Hongwei Xi} {and} \bibinfo{person}{Frank
  Pfenning}.} \bibinfo{year}{1999}\natexlab{}.
\newblock \showarticletitle{Dependent Types in Practical Programming}. In
  \bibinfo{booktitle}{\emph{Proceedings of the 26th Symposium on Principles of
  Programming Languages}} \emph{(\bibinfo{series}{POPL '99})}.
  \bibinfo{pages}{214--227}.
\newblock


\end{thebibliography}


\end{document}